\documentclass[]{article}

\usepackage{amsmath, amsthm, amssymb, amsfonts}
\usepackage{mathtools}
\usepackage{hyperref}
\usepackage{graphicx}
\usepackage{dsfont}
\usepackage{fullpage}
\usepackage{color}
\usepackage{soul} 
\usepackage[title]{appendix}
\usepackage{tikz}
\usetikzlibrary{patterns}
\usetikzlibrary{shapes.multipart}
\usetikzlibrary{arrows}

\theoremstyle{definition}

\definecolor{labelkey}{cmyk}{.4,.2,0,0}

\newcommand{\be}{\begin{equation}}
\newcommand{\ee}{\end{equation}}
\newcommand{\bea}{\begin{eqnarray}}
\newcommand{\eea}{\end{eqnarray}}
\newcommand{\nn}{\nonumber}

\newcommand{\inlaw}{\overset{\rm in \, law}{=}}

\usepackage{titlesec}
\titleformat{\section}{\large\bf}{\thesection}{1em}{}
\titleformat{\subsection}[runin]{\bf}{\thesubsection}{1em}{}[.]

\usepackage{authblk}
\title{This is some thing}
\author[1*]{Pierre Le Doussal}
\affil[1]{\normalsize Laboratoire de Physique de l'\'Ecole Normale Sup\'erieure, ENS, Universit\'e PSL, CNRS, Sorbonne Universit\'e, Universit\'e de Paris, 75005 Paris, France}

\title{\bf \large Large deviations of the largest eigenvalue for deformed GOE/GUE random matrices via replica}
\date{}

\begin{document}

\maketitle

\begin{abstract}
We study the probability distribution function $P(\lambda)$ 
of the largest eigenvalue $\lambda_{\rm max}$ of $N \times N$ random matrices of the form $H + V$, where $H$ belongs to the GOE/GUE ensemble and $V$ is a full rank deterministic diagonal perturbation. This model is related
to spherical spin glasses and semi-discrete directed polymers. In the large $N$ limit,
using the replica method introduced in Ref. \cite{TrivializationUs2014}, we obtain the rate function 
${\cal L}(\lambda)$ which describes the upper large deviation tail 
$P(\lambda) \sim e^{- \beta N {\cal L}(\lambda) }$. We also obtain the cumulant generating function
$\langle e^{N s {\lambda}_{\max} } \rangle \sim e^{N \phi(s)}$ and the overlap of the optimal eigenvector
with the perturbation $V$. For suitable $V$, a transition generically occurs in the rate functions. For the GUE it has a direct interpretation as
a localisation transition for tilted directed polymers with competing columnar and point disorder. 
Although in a different form, our results are consistent
with those obtained recently by McKenna in
\cite{McKenna2021}. Finally, we consider briefly the quadratic optimisation problem in presence 
of an additional random field and obtain its large deviation rate function, 
although only within the replica symmetric 
phase. 
\end{abstract}

\bigskip

{\it $^*$Corresponding author: pledoussal@yahoo.fr}

\section{Introduction}

\subsection{Deformed Wigner matrices: overview} 

In this paper we consider deformed Wigner matrices $M$ of size $N \times N$ of the form 
\be  \label{deformed} 
M = J H + V   \quad , \quad H \in \begin{cases} \text{GOE}(N) \quad \beta =1  \\
\text{GUE}(N) \quad \beta =2 \end{cases} \quad , \quad V = {\rm diag}(v_1,\cdots,v_N) \quad , \quad J>0
\ee 
We restrict to $H$ being either a real symmetric random matrix belonging to the Gaussian orthogonal ensemble 
(GOE) or an hermitian random matrix belonging to the Gaussian unitary ensemble 
(GUE) \cite{MehtaBook,ForresterBook} and $\beta$ is the Dyson index. We normalize $H$ so that its spectrum is a semi-circle of support
$[-2,2]$ in the large $N$ limit, which corresponds to choosing its probability distribution function (PDF) proportional to
${\cal P}(H) \sim e^{- \frac{\beta N}{4} {\rm Tr} H^2}$.
Here $J>0$ is a real parameter, and the perturbation $V$ is a deterministic real matrix chosen to be
diagonal. In this paper we will use the physics bracket notation for expectation values w.r.t.
$H$, i.e. $\langle O \rangle \equiv \mathbb{E}[O]$ 

These deformed matrices appear in many contexts, in physics, mathematics
and beyond \cite{EdwardsJones1976,BrezinHikami1996,BrezinHikami1997,BrezinHikamiBook,Johansson2001}.
A question of particular interest, on which we will focus here, is the 
statistics of the largest eigenvalue of the random matrix $M$, $\lambda=\lambda_{\rm max}(M)$.
Let us recall only a few notable examples, among all the problems where these deformed matrices appear.

In May's problem of ecosystems \cite{May1972}
one asks about the stability near an equilibrium for $N$ species of populations $x_i$
evolving as $\dot x_i = - \mu  x_i + H_{ij} x_j$, where the matrix $H_{ij}$ belongs to the GOE and 
models the random interactions, $\mu$ is a control parameter. 
The system is thus stable if $\mu > \lambda_{\rm max}(H) $. The deformed case 
thus corresponds to May's problem in
presence of (here deterministic) inhomogeneities 
\be 
\dot x_i = v_i  x_i + H_{ij} x_j   \label{May} 
\ee
and the system is stable iff $\lambda_{\rm max}(M)<0$. This heterogeneous stability problem was 
studied recently in \cite{MergnyMajumdar}, with explicit results in the
case where the $v_i$ form an equally spaced lattice. 
Note that the full time evolution was studied in \cite{Schomerus2016} in the homogeneous case.

The deformed GOE/GUE also appear in the context of the Dyson Brownian motion (DBM)
\cite{MehtaBook,ForresterBook}. Consider a matrix $W(t)$ whose entries perform real/complex Brownian motions, which are independent, up to the constraint that $W$ is real symmetric/hermitian. More precisely,
\be 
dW_{ii}=\sqrt{\frac{2}{\beta N}} \, dB_{ii} \quad , \quad  dW_{ij}=\sqrt{\frac{1}{\beta N}} \, \begin{cases} 
dB_{ij} \\
dB_{ij} + i dB'_{ij} \end{cases} \quad i < j
\ee
where $B_{ii}$, $1 \leq i \leq N$, and $B_{ij},B'_{ij}$, $1 \leq i<j \leq j$, are a collection of independent standard Brownian motions, and $\beta=1$ for the GOE, $\beta=2$ for the GUE.
If the chosen initial condition is $W(0)=V$, then the matrix $W(t)$ at a fixed time $t$ 
has the same distribution as $M$ in \eqref{deformed}, with $J=\sqrt{t}$. Its eigenvalues
$\lambda_j(t)$ perform a DBM, i.e  are solutions of the stochastic evolution
\be 
\dot \lambda_i(t) = \frac{1}{N} \sum_{j \neq i} \frac{1}{\lambda_j(t)-\lambda_i(t)} + \sqrt{\frac{2}{\beta N}} db_i(t) 
\ee 
with initial condition $\lambda_j(0)=v_j$, where the $db_i(t)$ are $N$ independent standard Brownian motions.

A third example is a directed polymer in dimension $1+1$ in presence 
of both point disorder and columnar disorder. It is the semi-discrete version known as the O'Connell-Yor 
polymer \cite{OYPolymer}. The polymer can live only along $N$ parallel horizontal columns, and can jump only upward from one to the next.
Its path $y(\tau)$, $\tau \in [0,L]$, $y \in {1,2,\cdots,N}$, is an increasing function which jumps 
by $+1$ at positions $\tau_j$ from the $j$-th column to the $j+1$-th, 
with $\tau_0=0 \leq \tau_1 \leq \cdots \leq \tau_{N-1}  \leq \tau_N=L$. The energy collected by the polymer along its path is the sum 
of (i) $\sum_{j=1}^N  (B_j(\tau_j)-B_j(\tau_{j-1}))$ where $B_j(\tau)$ are $N$ independent Brownian motions, modeling
point disorder on each column, 
and of (ii) $\sum_{j=1}^N v_j \ell_j$ with $\ell_j=\tau_j-\tau_{j-1}$ modeling columnar disorder. Then it is known that the energy 
$E^{\rm poly}_{\max}$ of the path with maximal energy is distributed as the largest eigenvalue of
a deformed random matrix, more precisely (see 
e.g. \cite{Baryshnikov,Gravner,Doumerc,Houdre,OConnell1,OConnell2,OConnellYor,Bougerol})
\be 
E^{\rm poly}_{\max}  \,  \inlaw \, 
 \, L \, \lambda_{\rm max} ( M=V + J \, H ) 
 \quad , \quad J= \sqrt{\frac{N}{L}} 
\ee
where $H$ belongs to the GUE(N). Here $J^2=N/L$ measures the inclination of the polymer path 
with respect the direction of the columns. It is
kept fixed as both $N$ and $L$ becomes large. This model was revisited
recently in \cite{Krajenbrink2021}, and we will compare our results with this work. 
As pointed out there, there is also a relation between 
the the $j$-th component $\psi_1(j)$
of the eigenvector $\psi_1$ associated to the largest eigenvalue of $M$
and a polymer observable, i.e. the total length $\ell_j^0=\tau^0_j-\tau^0_{j-1}$ of the optimal polymer 
along column $j$. It holds only on average and one has, for $j=1,\dots,N$
\be  \label{relationBH} 
\mathbb{E}_B( \ell_j^0 ) = L \, \langle  (|\psi_1(j)|^2) \rangle
\ee 
where $\psi_1$ is normalized to unity, the expectation values in the l.h.s. is w.r.t. the Brownians $B_j$
and on the r.h.s. w.r.t. the GUE matrix $H$ (at fixed $V$). 

Finally, GOE random matrices also appear in the study of the spherical spin glass
\cite{KosterlitzSpherical1976,CrisantiSommers1992}. The model is
defined by the energy function ${\cal H}[{\bf x}] = J \sum_{i,j=1}^N H_{ij} x_i x_j$ on the
sphere $\sum_i x_i^2 = N$. In that context the deformed model $J H \to M=J H + V$
corresponds to a spherical spin glass with anisotropies. Since the PDF of $H$ is
rotationally invariant, a rank one
perturbation is equivalent to an additional ferromagnetic coupling
$J_0 \sum_{i,j=1}^N x_i x_j$ with $J_0=v_1/N$. The ground state
energy $E_0$ of the spin glass model, , i.e. the maximum of ${\cal H}[{\bf x}]$ on the sphere, is then given by the  largest eigenvalue of $M$, i.e. one has $E_0=\lambda_{\rm max}(M)$.
The typical fluctuations of $E_0$, and more generally of the free energy of the
model at finite temperature, have been studied recently, 
mostly for rank one perturbations or in presence of an external field
\cite{BailkLeeTW2016,BailkLeeFerro2017,BaikPLD2021}.
Note that another ensemble, Wishart matrices, appear in the bipartite spin glass
problem \cite{BailkLeeBipartite2020}. 
\\

The main question of interest here is the behavior of the largest eigenvalue of $M$
and how it depends on the perturbation $V$. A much studied situation is the case where $V$ has rank one, also called a spike, $V={\rm diag}(v_1,0,\cdots,0)$. 
In that case the famous Baik-Ben-Arous-P\'ech\'e (BBP) transition 
\cite{EdwardsJones1976,BBP2005,BBPPeche2006} occurs as $N \to +\infty$: 
upon increasing the strength of the perturbation $v_1$, a single eigenvalue (the largest one) becomes an
outlier and
detaches from the Wigner sea, namely 
\be \label{BBP1} 
v_1 < J \quad : \quad \lambda_{\max}(M) = 2 J  \quad , \quad v_1 > J \quad : \quad \lambda_{\max}(M) = v_1 + \frac{J^2}{v_1} 
\ee
In the polymer language, the BBP transition corresponds to a transition of localization of the polymer on the best column, meaning that for $v_1>J$  the optimal polymer path remains on that column for a finite fraction of its length $L$.
More precisely, the occupation length of the column $1$ is $\ell_1^{\rm opt} = L (1- \frac{J^2}{v_1^2})_+$
\cite{BBP2005,Krajenbrink2021}.
Eq. \eqref{BBP1} are leading estimates, and the subleading typical fluctuations of $\lambda_{\max}(M)$ 
have been much studied in various contexts \cite{KnowlesYin2011}.
They exhibit a transition from being $O(N^{-2/3})$ with a Tracy-Widom distribution for $v_1<J$ 
to being $O(N^{-1/2})$ with a Gaussian distribution for $v_1>J$ (for GOE/GUE, but in that case the distribution
is not universal with respect to the distribution of the entries \cite{CapitaineSpikeNonUniversal2009}). The critical fluctuation regime, called
BBP$_1$, occurs when $v_1-J = O(N^{-1/3})$ and is non-trivial. In the case of a perturbation $V$ of finite rank $n>1$, a finite sequence 
of such transitions occurs \cite{BBPPeche2006} as the second, third, and so-on eigenvalues become outliers. These transitions
are independent unless the $v_j$ are nearly degenerate, e.g. for $n=2$ and $v_1-v_2=O(N^{-1/3})$, leading to
a BBP$_2$ transition and so on (see however \cite{KnowlesYin2014}). 
\\

Another interesting situation, on which we will focus here, is the case where $V$ is of rank $O(N)$, i.e. a full rank perturbation. One defines
the empirical eigenvalue density $\rho_N(v)$ of the matrix $V$, and one assumes 
that it has 
a good and smooth limit at large $N$
\be 
\rho_N(v) = \frac{1}{N} \sum_i \delta(v-v_i) \quad , \quad \rho(v) = \lim_{N \to +\infty} \rho_N(v) 
\ee 
Here we will further assume that $\rho(v)$ has a bounded support with a right (upper) edge at $v=v_e$. 
In the large $N$ limit the density of eigenvalues of the matrix $J H$ is the semi-circle
\be 
\rho_{sc}(\lambda) = \frac{1}{2 \pi J^2} \sqrt{(4 J^2 - \lambda^2)_+} 
\ee 
We will denote $\nu(\lambda)$ the eigenvalue density for the matrix $M= J H + V$ in the large $N$ limit. It is known
that it is given by the rules of free addition $\nu = \rho_{sc} \boxplus \rho$
\cite{Pastur1972,Voiculescu1991,Biane} (see \cite{CapitaineSpike2011} in the
context of spiked deformations). Introducing the Stieltjes transforms associated to
$V$ and $M$, they
obey
\be 
 G_M(y) = G_V(y- J^2 G_M(y)) \quad , \quad G_V(z)= \int dv \frac{\rho(v)}{z-v} 
 \quad , \quad G_M(y)= \int d\lambda \frac{\nu(\lambda)}{y-\lambda} 
\ee 
This allows to determine the right edge of the support of the density $\nu$ of $M$, i.e
the typical value of $\lambda_{\max}(M)$ (see below). The typical fluctuations of the largest eigenvalue 
in the case of the full rank perturbation have been less studied \cite{BorodinPeche2008}. 
Most asymptotic results apply to what we would call
the delocalized phase for the polymer, where the fluctuations 
are of the standard Tracy Widom type \cite{BorodinPeche2008,Shcherbina2011,BaikNadakuditi2014,CapitainePeche2015,LeeSchnelli2015}.
Recently we showed \cite{Krajenbrink2021} that for densities $\rho(v)$ which 
behave as $\rho(v) \simeq (v_e-v)^\alpha$ with $\alpha>1$ near
their edge, a polymer localization transition occurs (akin to a freezing transition). 
Around the transition $\lambda_{\max}(M)$ exhibits a novel critical fluctuation regime, which we characterized \cite{Krajenbrink2021}. 
\\

Finally note that the typical fluctuations in the case where $V$ is itself random, e.g. the $v_i$ are i.i.d., have also been 
considered \cite{JohanssonCrossover2007,LeeSchnelli2016,LeeSchnelli2013}. 
\\

{\bf Important note:} Every statement below about localization is only about {\it polymer localization}. Although \eqref{relationBH} 
gives a bridge with eigenvector localization, we will not study that aspect in detail here.

\subsection{Large deviations and aim of the paper} 

In this paper we will be interested in the large deviations for $\lambda=\lambda_{\max}(M)$ with
$M$ of the form \eqref{deformed} with $V$ a  full rank deterministic perturbation.
In the case of a rank one perturbation the large deviation function was obtained in \cite{Maida2007}, 
and in \cite{Donati2012} in the context of hermitian Brownian motion. For large $N$ and for $\lambda \geq 2$ the tail of the PDF of $\lambda=\lambda_{\max}(M)$ behaves as
\be \label{LD0} 
P(\lambda) \sim e^{- \beta N {\cal L}(\lambda) } 
\ee 
where explicit expressions for ${\cal L}(\lambda)$ were given there. The case of finite fixed rank $n$ was adressed in \cite{Benaych2012}. 
For a rank one perturbation, the joint large deviations of the largest eigenvalue and its eigenvector
was obtained in \cite{BiroliGuionnet2019}.

The large deviations in the full rank case were investigated only very recently by McKenna \cite{McKenna2021}.
A large deviation principle of a form analogous to \eqref{LD0} was proved, with a rate function obtained there.

The aim of the present paper is to calculate this large deviation rate function in a physicist's way,
using a quite different and non-rigorous replica method. This method was introduced in \cite{TrivializationUs2014}
to calculate the large deviations for the ground state energy of a spherical spin glass in a random field,
equivalently for the maximum of a random quadratic plus linear form, two problems also considered in e.g. \cite{DemboZeitouni2015,BaikPLD2021}.
We will mainly use this method to study the case of the deformed random matrix $M$ in \eqref{deformed}.
In the last part of the paper, however, we will extend our results to study the maximum of a quadratic form
involving $M$, plus a linear random field. In both cases
we will restrict to the replica symmetric (RS) phase. While for the first
problem this is likely not a restriction, for the second it was found
that already for $V=0$ there is a (small) region of the parameters
where the RS solution is invalid \cite{DemboZeitouni2015},
and one must consider instead a replica symmetry breaking saddle point
\cite{LacroixRSB}. With this caveat in mind, we present
our results for that second problem, leaving the 
complete solution for future work.  
\\

For completeness let us mention some recent works on related topics. 
In physics there were some works using replica to study deformed matrices, focusing
on different questions, either about the typical behavior of $\lambda_{\rm max}(M)$ \cite{Ikeda2023,Urbani2022},
or about the mesoscopic fluctuations of the eigenvalues in the bulk of $M$ \cite{GregRPmodel22}.

In the math literature, the large deviations of the largest (or the smallest) eigenvalue 
of random matrices were studied recently in various contexts, such as
generalized sample covariance matrices \cite{HussonMcKenna2024}, 
matrices with a variance profile \cite{Husson2022VarianceProfile,Ducatez2024VarianceProfile},
sharp subgaussian matrices \cite{GuionnetHusson2020}, free convolution models \cite{GuionnetMaidaFreeConvolution}. 
The sensitivity of the rate function to the tail of the distribution of entries for Wigner matrices 
was studied in \cite{Augeri}. For sharp sub-Gaussian distributions 
it matches the rate function of the GOE 
\cite{BordenaveLargestSharpSubGauss2014,AugeriLargestSharpSubGauss2021},
while in more general case it may exhibit an eigenvector localization 
 \cite{CookDucatezLargestSubGaussLoc2023}.
 
 Finally, while this work was in the last stage of completion, a paper
appeared in mathematics \cite{BoursierGuionnet2024}, which
also studies a full rank deformation of a GOE matrix, but in presence of an additional outlier,
encompassing the results of both \cite{McKenna2021} and \cite{Maida2007}.

\section{Main results} \label{sec:main} 

We consider the matrix $M = J H + V$ where $H$ is a GOE/GUE random matrix, $V$ is a diagonal deterministic matrix
of spectral density $\rho(v)$ in the large $N$ limit. We assume that it has a bounded support with an upper edge at $v=v_e$.
We study
$\lambda= \lambda_{\rm max}(M)$, which, as we recall, can also be seen as the optimal energy of a polymer, as 
well as the ground state energy of some spherical spin-glass model.

As $J$ is varied there are two possible phases, and a transition which occurs at $J=J_c$
with 
\be  \label{Jceq} 
J_c^2 = \frac{1}{\int dv  \frac{\rho(v) }{(v_e - v  )^2}} 
\ee 
For $J>J_c$ the optimal polymer path is delocalized, while it is localized for $J<J_c$. 
In the case where 
\be 
 \int dv  \frac{\rho(v) }{(v_e - v  )^2} = + \infty
\ee 
then $J_c=0$ and there is only a polymer delocalized phase. 
\\

\subsection{First result: Cumulant generating function (CGF)}

Using the replica method we first obtain, for real $s \geq 0$ and as $N \to +\infty$ 
\be \label{defphi1} 
 \langle e^{N s {\lambda}_{\max}(M) } \rangle \sim e^{N \phi(s)} 
\ee
The function $\phi(s)$, which we call the scaled CGF, 
must be convex, and has the following expressions. 
\\

(i) In the delocalized phase $J>J_c$, the the scaled CGF $\phi(s)$ has two expressions
\\

a) For $0 \leq s \leq s^*=s^*(J)$ it is given by 
\be \label{Phideloc0} 
 \phi(s) = \phi(s, {\sf z}_+(s)) \quad , \quad  \phi(s, z)  = s z - \frac{J^2}{\beta}  s^2 
- \frac{\beta}{2} \int dv \rho(v)  \log\left( \frac{z - v - 2 s J^2/\beta}{z - v} \right)  
\ee
where ${\sf z}_+(s)$ is the unique solution of $\partial_{z} \phi(s, z) |_{z={\sf z}_+(s)} = 0$
with ${\sf z}_+(s) \in [v_e + 2 s J^2/\beta,+\infty[$, explicitly of 
\be 
1 = J^2 \int dv  \frac{\rho(v) }{(z -  v  ) (z - v - 2 s J^2/\beta)} 
\ee

b) For $s \geq s^*$ it is given by 
\be \label{locreg} 
\phi(s)  = \phi(s,v_e + 2 s J^2/\beta) =  v_e s + \frac{J^2}{\beta}  s^2 
- \frac{\beta}{2} \int dv \rho(v)  \log\left( \frac{v_e - v}{v_e - v + 2 s J^2/\beta} \right)  
\ee 
The value of $s^*=s^*(J)$ is the positive root of the equation
\be \label{defsstar} 
1 = J^2 \int dv  \frac{\rho(v) }{(v_e -  v  + 2 s^* J^2/\beta ) (v_e - v )} 
\ee 
Equivalently, as $s$ increases, $s=s^*$ is the point at which ${\sf z}_+(s) - 2 s J^2/\beta$ reaches $v_e$, i.e. it is also
solution of
\be 
{\sf z}_+(s^*) = v_e + 2 s^* J^2/\beta
\ee 
Upon comparing \eqref{locreg} and \eqref{Phideloc0} one could
consider that for $s \geq s^*$, ${\sf z}_+(s)$ remains frozen to the ($s$-dependent) 
value $v_e + 2 s J^2/\beta$. 
\\

(ii) In the localized phase $J<J_c$, one has formally $s^*=0$ and for all $s \geq 0$, $\phi(s)$ has the same expression
as case (i)-(b), namely $\phi(s)  = \phi(s,v_e + 2 s J^2/\beta)$ as given by \eqref{locreg}.
\\

Thus, within the large deviations region $s \geq 0$, we can call "delocalized regime" the case $s \leq s^*$ and "localized regime" (or frozen) the case $s \geq s^*$. In terms of the polymer, they are dominated respectively by delocalized and partially localized polymer configurations, very much as in the typical case $s=0$
\cite{Krajenbrink2021}. To distinguish it from typical however we use the terminology "regime".\\

Note that ${\sf z}_+(s)$ is denoted $\kappa$ in the text below. Finally, 
the convexity of $\phi(s)$ is checked explicitly in Appendix \ref{app:convex}.

\subsection{Second result: Large deviation rate function}

The above result for the Laplace transform is consistent with the large deviation form for the PDF
of $\lambda= \lambda_{\rm max}(M)$,
\be \label{LD1} 
P(\lambda) \sim e^{- \beta N {\cal L}(\lambda) } 
\ee 
where the rate function ${\cal L}(\lambda)$ can be obtained from the Legendre transform
\be \label{Leg0} 
\beta {\cal L}(\lambda) = \max_{s \geq 0} ( s \lambda -  \phi(s)) 
\ee 
which 
leads to the pair of equation $s = \beta {\cal L}'(\lambda)$, $\lambda = \phi'(s)$. In particular
the typical value $\lambda_{\rm typ}$ of $\lambda_{\rm max}(M)$ is determined
by ${\cal L}'(\lambda_{\rm typ}) =0$ and thus corresponds to $s=0$ (although 
this is true in most cases, the most general criterion is ${\cal L}(\lambda_{\rm typ}) =0$).
The 
forms \eqref{LD1} and \eqref{Leg0} are valid for $\lambda \geq \lambda_{\rm typ}$,
which corresponds to $s \geq 0$.
\\

The result for ${\cal L}(\lambda)$ can be formulated in two stages as follows. 
\\

\begin{figure}[t]
   \includegraphics[width=0.33\columnwidth]{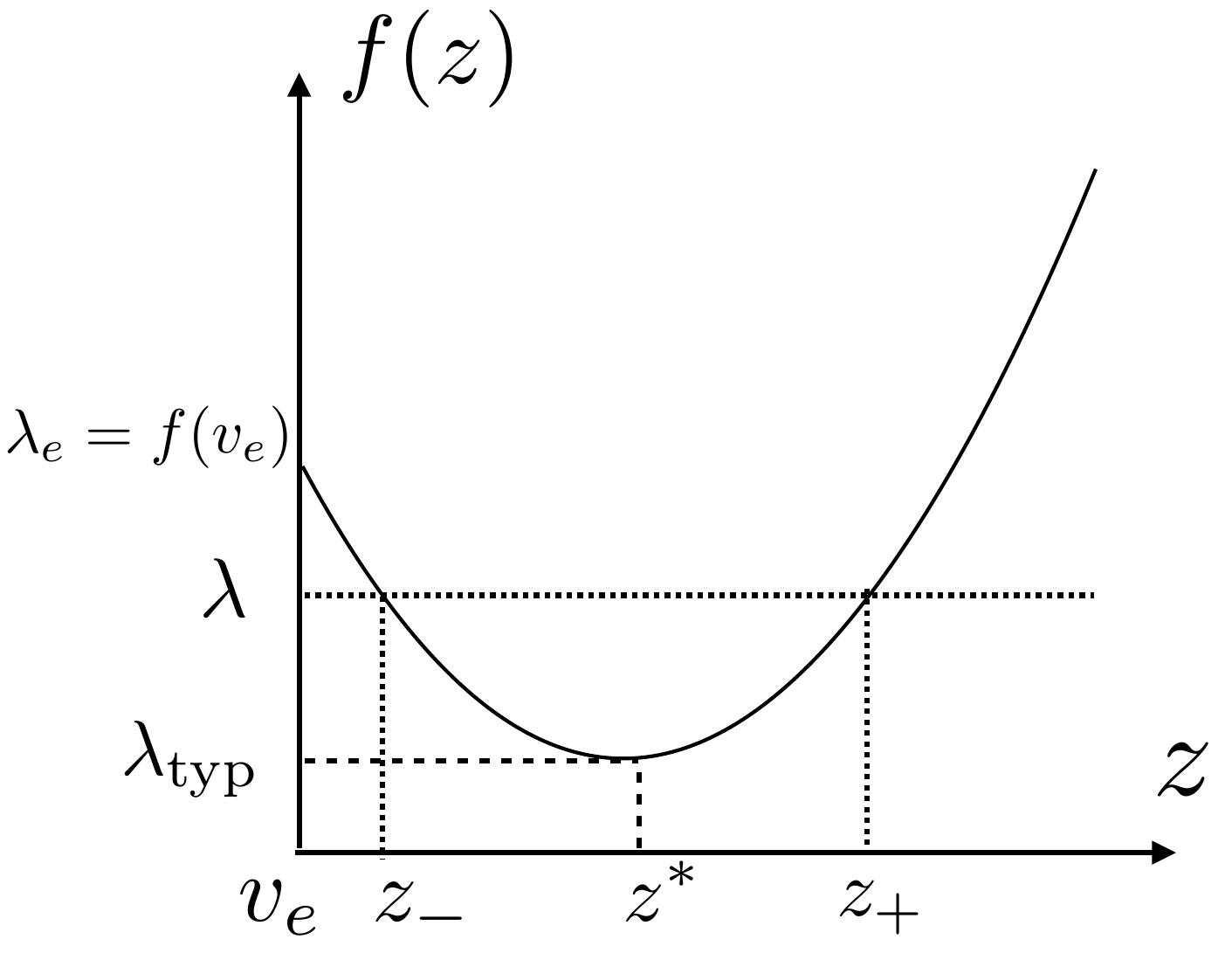}
    \includegraphics[width=0.33\columnwidth]{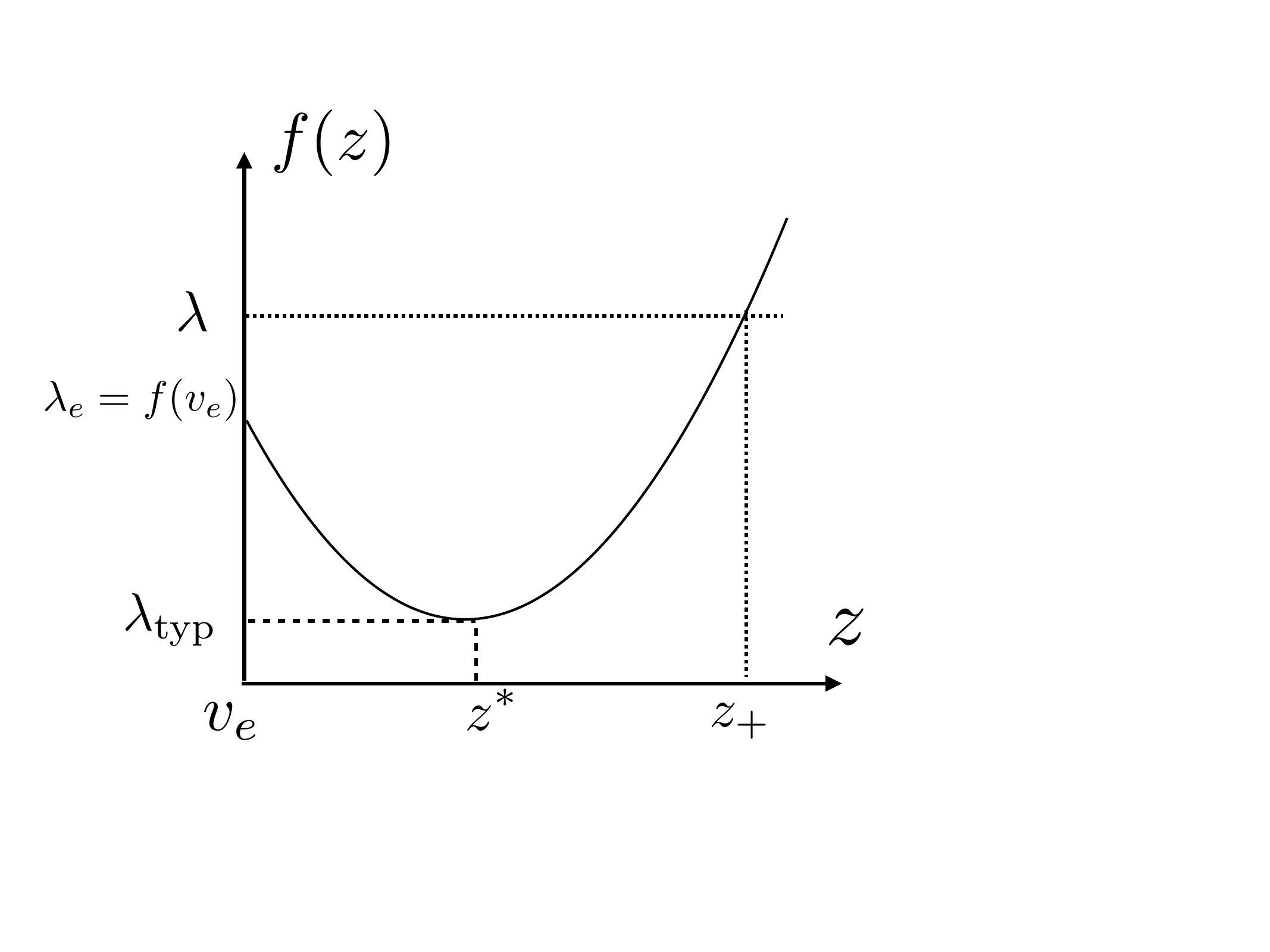}
     \includegraphics[width=0.33\columnwidth]{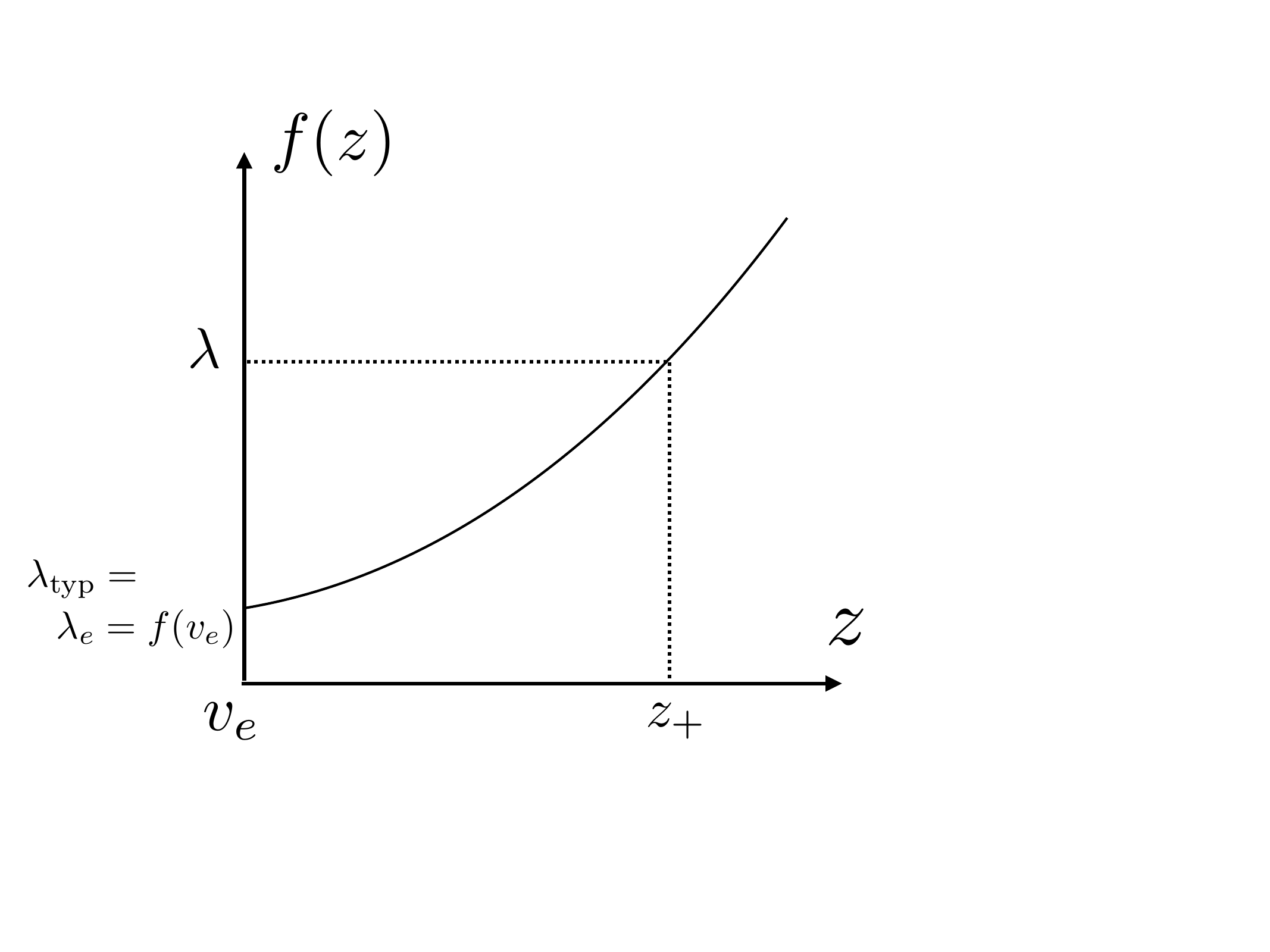}
   \caption{Roots $z_\pm$ of $\lambda=f(z)$ in the three different cases (i) left: delocalized phase $J>J_c$ and delocalized regime
   (ii) center: delocalized phase $J>J_c$ and localized regime
   (iii) right: localized phase $J<J_c$. In the delocalized phase $\lambda_{\rm typ}=f(z^*)$ where $z^*$ is
   the unique minimum of $f(z)$ in the interval $z \in [v_e,+\infty[$. In the localized phase $\lambda_{\rm typ}=\lambda_e=f(v_e)$.}
  \label{fig:Sketch}
\end{figure}

First, introduce the following function $f(z)$ of the real variable $z \in [v_e,+\infty[$
\be \label{deff} 
 f(z) := z + J^2 \int dv \frac{\rho(v)}{z-v} 
\ee 
The function $f(z)$ is strictly convex, i.e. $f''(z) >0$. It has a unique minimum on the interval $[v_e,+\infty[$.
Let us call it $z^*=z^*(J)$. It can be either inside the interval $]v_e,+\infty[$ or
at the boundary of this interval, see Fig. \ref{fig:Sketch}. Hence $z^*$ is determined
in either cases as
\bea
&& J > J_c \quad , \quad 1 = J^2 \int dv \frac{\rho(v)}{(z^*-v)^2} \quad , \quad z^* > v_e \\
&& J < J_c \quad , \quad z^* = v_e 
\eea
These two cases correspond respectively to the delocalized and the localized phase of the polymer. 
At the transition between the two phases (when it exists) one has $z^*(J_c)=v_e$. 
\\

Now, for a given value of $\lambda= \lambda_{\rm max}(M)$ consider the equation
\be \label{eqf} 
\lambda = f(z) 
\ee 
It can have roots on $[v_e,+\infty[$ only for $\lambda \geq f(z^*)$. The value of $z^*$ determines the {\it typical value} 
$\lambda_{\rm typ}$ of $\lambda= \lambda_{\rm max}(M)$. Indeed one has
\be \label{typical} 
\lambda_{\rm typ} = f(z^*)
\ee 
Thus in the localized phase $J<J_c$ the typical value is frozen at the ($J$ dependent) value 
$\lambda_{\rm typ} =  \lambda_e := f(v_e)$.
\\

This result for the typical value of the largest eigenvalue is consistent with the one obtained by quite different methods in Section III.A of \cite{Krajenbrink2021} (i.e. Eq. (81) there, with $ \mu= \lambda/J^2$ and $\theta=1/J^2$). 
It also comes from the fact that $z^*$ is the edge of the support of the measure 
$\rho_{sc} \boxplus \rho$ under free addition, see e.g. discussion in \cite{MergnyMajumdar}. 
The result in the localized phase is consistent with 
the discussion in \cite{Claeys} and in Section IV. C of \cite{Krajenbrink2021}.
\\



Consider now $\lambda \geq \lambda_{\rm typ}$. Consider first the delocalized
phase $J>J_c$. Depending on $\lambda$ the equation \eqref{eqf} on the interval $[v_e,+\infty[$ 
can have either two roots, on either side of $z^*$,
or a single root, to the right of $z^*$. This is illustrated in Fig. \ref{fig:Sketch}.
For
\bea
&& \lambda_{\rm typ} < \lambda \leq \lambda_e = f(v_e) = v_e + J^2 \int dv \frac{\rho(v)}{z-v_e} 
\quad , \quad \text{there are two roots $z_\pm$} \quad v_e < z_- < z^* < z_+ \\
&& ~~~~~~~\lambda > \lambda_e  \quad , \quad ~~~~~~~~~~~~~~~~~~~~~~~~~~~~~~~~~~~~~~~~~~~
\text{there is a single root $z_+$} \quad v_e  < z^* < z_+
\eea
These roots are functions of $\lambda$ and $J$, we will note $z_\pm=z_\pm(\lambda)$. While 
$z_+$ is an increasing and continuous function of $\lambda$, for $\lambda > \lambda_e$ the
smaller root $z_-$ disappears. The case $\lambda_{\rm typ} < \lambda < \lambda_e$
is called delocalized regime, and the case $\lambda > \lambda_e$ 
is called the localized regime. Note that in the case $f(v_e)=+\infty$ there is only the first 
(delocalized) regime with two roots. Conversely, in the localized phase $J<J_c$, one has 
$\lambda_{\rm typ}=\lambda_e$ and there is always a single root $z_+$
so one is always in the second (localized) regime.
\\

We can now give the result for the rate function ${\cal L}(\lambda)$. Let us first consider
the delocalized phase $J>J_c$ and regime, that is $\lambda_{\rm typ} \leq \lambda < \lambda_e$. 
In performing the Legendre transform \eqref{Leg0} it turns out that the variable $s$ corresponds 
to $2 J^2 s/\beta = z_+(\lambda) - z_-(\lambda)$ defined above. One can then replace it
in $\beta {\cal L}(\lambda) = s \lambda -  \phi(s)$ and obtain
\bea 
&& {\cal L}(\lambda)   =   \frac{1}{4 J^2} ( z_+-z_-) (2 \lambda - z_+ - z_-)  
+  \frac{1}{2} \int dv  \rho(v) \log \frac{z_- -v}{z_+ - v} \label{ratedeloc0} 
\eea 
where $z_\pm=z_\pm(\lambda)$ are the two roots of \eqref{eqf}-\eqref{deff}. 
\\

In the localized regime of the delocalized phase, $J>J_c$ and $\lambda \geq \lambda_e$, and in the
localized phase $J<J_c$ one obtains
\be \label{Ellloc} 
 {\cal L}(\lambda)  =  \frac{1}{4 J^2} ( z_+-v_e) (2 \lambda - z_+ - v_e)  
+ \frac{1}{2} \int dv \rho(v)  \log\left( \frac{v_e - v}{z_+ - v} \right)  
\ee
where $z_+=z_+(\lambda)$ is the single root of \eqref{eqf}-\eqref{deff} with $z_+ > v_e$. 
This amounts to freeze and replace $z_- \to v_e$ in  \eqref{ratedeloc0}.
\\

For practical calculations it is actually easier to compute ${\cal L}(\lambda)$ in the delocalized phase as
\bea 
&& {\cal L}(\lambda) = \int_{\lambda_{\rm typ}}^\lambda dx {\cal L}'(x) 
= \int_{\lambda_{\rm typ}}^\lambda dx \frac{z_+(x)- z_-(x)}{2 J^2}  \quad , \quad \lambda_{\rm typ} \leq \lambda \leq  \lambda_e \quad , \quad
\text{delocalized regime}  \\
&& {\cal L}(\lambda) = {\cal L}(\lambda_e) + \int_{\lambda_{e}}^\lambda dx {\cal L}'(x) 
= \int_{\lambda_{\rm typ}}^\lambda dx \frac{z_+(x)- v_e}{2 J^2}  \quad , \quad  \lambda \geq  \lambda_e \quad , \quad
\text{localized regime} 
\eea 
In the localized phase one uses the second expression with ${\cal L}(\lambda_e) =0$. 

{\bf Remark}. Comparing with the previous section, note that ${\sf z}_+(s)=z_+(\lambda)$ are the
same quantity (the largest root $z_+$ 
of \eqref{eqf}) when 
$\lambda$ and $s$ are related by the Legendre transform.
\\

{\bf Behavior of the rate function near the typical value}. 
Having obtained the large deviation rate function
${\cal L}(\lambda)$, we can
examine its behavior near the typical value $\lambda_{\rm typ}$. This is performed in details
in Section \ref{subsec:behavior}. The results are as follows

- In the delocalized phase $J>J_c$ we find that it behaves with an exponent $3/2$, i.e.
${\cal L}(\lambda) \propto (\lambda - \lambda_{\rm typ})^{3/2}$. As shown there,
 this behavior precisely matches (including the coefficient) the stretched
 exponential upper tail of the Tracy-Widom distribution which describes instead the
 typical fluctuation $O(N^{-2/3})$ of $\lambda_{\rm max}(M)$ in
 the delocalized phase (these typical fluctuations were obtained 
 in \cite{Krajenbrink2021} for the present model). This extends to deformed random matrices
 a similar matching noted in \cite{TrivializationUs2014} in the case $V=0$. 

- In the localized phase $J<J_c$ the rate function has the more standard quadratic behavior, ${\cal L}(\lambda) \propto (\lambda - \lambda_{\rm typ})^{2}$ with now $\lambda_{\rm typ}=\lambda_e=f(v_e)$,
where $f(z)$ is defined in \eqref{deff}. From
the coefficient we obtain the variance (defined from the large deviation side)
\be \label{varianceloc0} 
{\rm Var} (\lambda-\lambda_e) = \frac{2 J^2}{\beta N} (1 - \frac{J^2}{J_c^2})
\ee 
which vanishes linearly at the transition.
The quadratic behavior is consistent with the upper tail of the
typical fluctuations which were obtained in \cite{Krajenbrink2021} and
found indeed to be Gaussian (curiously, up to a factor $2$ mismatch,
see Section \ref{subsec:behavior}). 

- The behavior around the transition point at $J_c$ interpolates between the two above
behaviors, and is more involved. It is worked out in 
Appendix \ref{app:interpolation}, and the result is displayed in Section \ref{subsec:behavior}. 
\\

{\bf Remark}.  
The result for ${\cal L}(\lambda)$ appears in a form a bit different from the one of McKenna
\cite{McKenna2021}. We have tested the equivalence
on some examples in Appendix \ref{app:McKenna}.
 The result for $\phi(s)$ was not given there. Nor was the localized phase
$J< J_c$ considered there.

\subsection{Occupation measure and overlap with the perturbation}

There is a simple interpretation of the above results. Let us denote ${\bf x}_{\max}$ the "optimal" eigenvector of $M$, i.e. the eigenvector associated to 
the largest eigenvalue $\lambda_{\rm max}(M)$  of $M$. It is
normalized as ${\bf x}_{\max}^T {\bf x}_{\max}=\sum_{i=1}^N |x_{\rm max,i}|^2 = N$.
One can define the conditional expectation 
\be \label{defoccupation} 
\rho_\lambda(v) = \left< \left(\frac{1}{N}  \sum_i \delta(v-v_i) |x_{\rm max,i}|^2 \, \bigg| \, \lambda_{\rm max}(M)= \lambda \right) \right>
\ee 
which is normalized to unity $\int dv \rho_\lambda(v)=1$ and vanishes for $v>v_e$. 
It contains information about the overlap of the optimal eigenvector with the perturbation,
defined as $\frac{1}{N} {\bf x}_{\max}^T V {\bf x}_{\max}$ (overlap here is meant
as in spin glasses, see e.g. \cite{BaikPLD2021}, and not as in non-Hermitian matrices \cite{ChalkerMehlig}).
In the GUE case its has also a direct interpretation in terms of the {\it occupation length} of the
optimal polymer configuration along the columnar defects
\cite{Krajenbrink2021}, see \eqref{relationBH} where ${x}_{\max,i}$ is denoted
$\psi_1(i)$, hence we call it the occupation measure.

We obtain the occupation measure in the limit $N \to +\infty$. In
the delocalized regime $J > J_c$, $\lambda_{\rm typ} \leq \lambda <\lambda_e$, it reads, 
\be 
\rho_\lambda(v) = J^2  \frac{\rho(v)}{ (z_+(\lambda) - v  )(z_-(\lambda) -  v)}  \quad , \quad v \leq v_e 
\ee 
where $z_+(\lambda) >z_-(\lambda) >v_e$ are the two roots of $\lambda=f(z)$ where $f(z)$ is
defined in \eqref{deff}. In the localized regime $\lambda> \lambda_e$ the occupation measure develops an
atomic part at the edge of the spectrum of $V$, and one has
\be 
\rho_\lambda(v) =  J^2  \frac{\rho(v)}{ (z_+(\lambda) - v  )(v_e -  v)}  + r_c \delta(v-v_e)  \quad , \quad 
r_c = \frac{\lambda-\lambda_e}{z_+(\lambda) - v_e} = 1 - J^2 \int dv \frac{\rho(v)}{ (z_+(\lambda) - v  )(v_e -  v)} 
\ee 
where $0<r_c<1$ such that $\rho_\lambda(v)$ is normalized to unity.
In the polymer problem, it corresponds to a regime where the optimal polymer
spends a fraction $r_c$ of its length on the optimal columnar defects with $v \simeq v_e$. 

The present result extends to the large deviation regime $\lambda>\lambda_{\rm typ}$
the result obtained in \cite{Krajenbrink2021} Section III.B for the typical fluctutations, i.e.
for $\lambda=\lambda_{\rm typ}$. In that case $z_+=z_-=z^*$ in the delocalized
regime which recovers (93) there, and $z_+=z_-=v_e$ in the localized regime,
which recovers (94) there, since in the limit $\lambda \to \lambda_e$ one has
$r_c \to 1 - \frac{J^2}{J_c^2}$.

\subsection{Results for quadratic optimization in presence of a random field}

We have also considered the same quadratic optimization problem in presence of an independent Gaussian random field. This extends our work in \cite{TrivializationUs2014} 
to the deformed GOE/GUE case. We consider the random variable 
\be  
\lambda_{\sigma} = \max_{{\bf x} \in \Omega_N} ( {\bf x}^\dagger (J H + V) {\bf x}  +   {\bf h}^\dagger {\bf x}  + {\bf x}^\dagger {\bf h}  ) 
\ee 
where ${\bf h}$ is a real/complex $N$ component Gaussian random field 
with variance $\langle h_i^* h_j \rangle= \beta \sigma^2 \delta_{ij}$. 
Since we will always consider expectation values both over $H$ and the random field, 
we use the same bracket symbol. For $\sigma=0$ one recovers the case
studied above and in this section we use the notations $\lambda_0 = \lambda_{\rm max}(M)$,
$\lambda^{\rm typ}_0 = \lambda_{\rm typ}$. 

As announced above, our results are valid only
in the region where the replica symmetric solution is valid.
We have not attempted to estimate the boundaries of this region. 
These are known in the case $V=0$ \cite{DemboZeitouni2015,LacroixRSB}
and are far from the region around the typical value (which 
is described by the replica symmetric solution). 

We show that the cumulant generating function for $s>0$, and 
the probability $P(\lambda) $ of $\lambda=\lambda_\sigma$ for 
$\lambda > \lambda^{\rm typ}_\sigma$ take the large deviation forms 
\be  \label{LD0RF} 
  \langle e^{N s {\lambda}_\sigma } \rangle \sim e^{N \phi_\sigma(s)} 
\quad , \quad  P(\lambda) \sim e^{- \beta N {\cal L}_\sigma(\lambda) } 
\ee

It turns out that the two functions $\phi_\sigma(s)$ and ${\cal L}_\sigma(\lambda)$
can be obtained from the ones for $\sigma=0$,
displayed in the previous subsections, and
which we will denote here for convenience 
$\phi_0(s) = \phi(s)$ and ${\cal L}_0(\lambda) = {\cal L}(\lambda) $.
One defines 
\be
\tilde J^2 := J^2 + \beta \sigma^2  
\ee 
Then the analog of the delocalized phase is now $\tilde J > J_c$ 
where $J_c$ is the zero field transition coupling given in \eqref{Jceq}. We now discuss
this phase. In that phase there is the analog of a delocalized regime
$0< s <s^*=s^*(J,\sigma)$ and the analog of a localized regime $s> s^*=s^*(J,\sigma)$.
The value of $s^*$ is simply the one in zero
field with the substitution $J \to \tilde J$, i.e. $s^*(J,\sigma)=s^*(\tilde J)$
where the function $s^*(J)$ was defined in \eqref{defsstar}. Next we find that 
\be 
\phi_\sigma(s) = \sigma^2 s^2 + \tilde \phi_0(s) \quad , \quad 
\tilde \phi_0(s) = \phi_0(s) |_{J^2 \to J^2+  \beta \sigma^2} 
\ee 
where $\phi_0(s)$ is the scaled CGF for $\sigma=0$,
which was given in \eqref{Phideloc0}. This property 
holds both for $s>s^*$ and $s<s^*$, where in the first case
$\phi_0(s)=\phi(s)$ is given by \eqref{Phideloc0} and
in the second case it is given by 
\eqref{locreg}. 

From it, the rate function for the probability 
can be obtained, from 
\be 
\beta {\cal L}_\sigma(\lambda) = \max_{s \geq 0} (\lambda s - \phi_\sigma(s)) 
= \max_{s \geq 0} (\lambda s - \sigma^2 s^2 - \tilde \phi_0(s))  
\ee 
First one finds that the typical
value is given by the same substitution
$\lambda_\sigma^{\rm typ}= \lambda_0^{\rm typ}|_{J \to \tilde J}= \lambda_{\rm typ}|_{J \to \tilde J}$. 
The rate function ${\cal L}_\sigma(\lambda)$ can also be obtained from ${\cal L}_0(\lambda)$ the rate function at zero field.
Indeed defining 
\be \label{corr} 
\tilde {\cal L}_0( \lambda) = {\cal L}_0(\lambda)|_{J^2 \to J^2+  \beta \sigma^2} 
\ee 
we obtain the parametric representation for the derivative ${\cal L}_\sigma'(\lambda)$, 
by elimination of $\tilde \lambda_0$ in 
the system
\bea
&& \lambda = \tilde \lambda_0 + 2 \sigma^2 \beta \tilde  {\cal L}_0'(\tilde \lambda_0) \label{sys1} \\
&&  {\cal L}_\sigma'(\lambda) =  \tilde {\cal L}_0'(\tilde \lambda_0) \label{sys2} 
\eea 
This leads to the following formula for the rate function itself
\bea 
&& {\cal L}_\sigma(\lambda)  = 
\tilde {\cal L}_0(\tilde \lambda_0[\lambda]) + \sigma^2 \beta \tilde {\cal L}_0'(\tilde \lambda_0[\lambda])^2
\eea 
where $\tilde \lambda_0[\lambda]$ is obtained by inversion of \eqref{sys1}. 
This equation is valid both for $\lambda < \lambda_\sigma$ (analog of delocalized
regime) and $\lambda > \lambda_\sigma$ (analog of localized
regime), where $\lambda_\sigma^e = [  f(v_e)  + 2 \sigma^2 s^*(J)]_{J \to \tilde J}$. 
In each regime one must use in \eqref{corr} a different expression for 
${\cal L}_0(\lambda)$, given respectively by 
\eqref{ratedeloc0} 
for $\lambda < \lambda_e$ (first regime) 
and by \eqref{Ellloc} for $\lambda > \lambda_e$ (second regime). 
In the analog of the localized phase only the second regime exists. 

\section{Replica calculation} \label{sec:replica}  

\subsection{Main idea of the calculation}

We now derive the above result, by a replica method similar to the one introduced in 
\cite{TrivializationUs2014}. To this aim we study the partition sum of the following spherical model
\be  \label{ZNT} 
Z_N(T) = \int_{{\bf x} \in \Omega_N}  d {\bf x} \, \delta( {\bf x}^\dagger {\bf x} - N) \, 
e^{ \frac{1}{T} {\bf x}^\dagger (J H + V) {\bf x}  }
\quad , \quad \Omega_N = \begin{cases}  \mathbb{R}^N (\beta=1) \\
 \mathbb{C}^N (\beta=2) \end{cases} 
\ee 
where we have introduced the temperature $T>0$.
Here ${\bf x}=(x_1,\dots,x_N)$ is a $N$-vector, either real or complex entries $x_i$, 
$V={\rm diag}(v_1,\dots,v_N)$ where the $v_i$ are deterministic real numbers, and
$H$ is symmetric ($\beta=1$) or hermitian ($\beta=2$) random matrix, drawn respectively
from the GOE(N) or the GUE(N) ensemble, i.e. with PDF $\sim e^{- \frac{\beta N}{4} {\rm Tr} H^2}$,
with the semi-circle mean spectral density at large $N$ with support $[-2,2]$. 
In Eq. \eqref{ZNT} $Z_N(T)$ can be interpreted for $\beta=1$ as the partition sum of a spherical spin glass model with random Gaussian couplings and in presence of anisotropies (with energy $\sum_{i=1}^N v_i x_i^2$). 

We are interested in the PDF of (minus) the intensive free energy \footnote{we use a different font not to confuse it 
with the function $f(z)$ defined in the introduction} ${\sf f}(T)$, and in particular its zero temperature limit
which allows to retrieve the largest eigenvalue of the matrix $M=J H + V$ (which has only real eigenvalues) as follows
\be
{\sf f} (T) =  \frac{T}{N} \log Z_N(T)     \quad , \quad \lim_{T \to 0} {\sf f}(T) = 
\max_{{\bf x} \in \Omega_N} {\bf x}^\dagger (J H + V) {\bf x} =
\lambda := \lambda_{\max}(M)
\ee 
since in the limit $T \to 0$ the integral in \eqref{ZNT} is dominated by the maximum of
the integrand.

The idea now is to study the integer moments 
$\langle Z_N(T)^n \rangle$ where the expectation is w.r.t. the GOE/GUE measure.
As in \cite{TrivializationUs2014} we compute the large $N$ asymptotics of these moments, using replica and a saddle point method,
which takes the form
\be   \label{asymptotics} 
\langle Z_N(T)^n \rangle \sim e^{N S_n(T)} 
\ee   
to leading order at large $N$, where $S_n(T)$ is the saddle point action.
From these moments one extracts, by first taking $n \to 0$ and second $T \to 0$, 
the mean value (which coincides with the typical value) of $\lambda_{\max}(M)$. 
Next one obtains the cumulant generating function 
and its scaled CGF $\phi(s)$ defined in
\eqref{defphi1}, through
\be \label{defphi2} 
\lim_{n = s T, T \to 0} \langle Z_N(T)^n \rangle = 
\lim_{T \to 0} \langle e^{N s \, {\sf f}(T)}  \rangle
= \langle e^{N s \, \lambda_{\max}(M) } \rangle \simeq g(s) e^{N \phi(s)} 
\ee
Hence $\phi(s)$ is obtained from the action at the saddle point $S_n(T)$,
formally $\phi(s) = \lim_{T \to 0} S_{n=s T}(T)$.
It requires (i) an analytical continuation in $n$ and then (ii) setting $n=s T$ with 
$s$ fixed as $T \to 0$. The analytical continuation in $n$ will be possible by using the
replica-symmetric ansatz which appears to be the correct one in these types of random
matrix problems. 

\subsection{Replica saddle point equations} 
\label{subsec:replicaSP} 

We are interested in the integer moments
\be  \label{ZN1} 
\langle Z_N(T)^n \rangle = \prod_{a=1}^n
\int_{{\bf x}_a \in \Omega_N}   d {\bf x}_a 
 \, \delta( {\bf x}_a^\dagger {\bf x}_a - N) \, \langle  e^{ \frac{1}{T} \sum_a {\bf x}_a^\dagger (J H + V) {\bf x}_a } 
  \rangle  \quad , \quad \Omega_N = \begin{cases}  \mathbb{R}^N (\beta=1) \\
 \mathbb{C}^N (\beta=2) \end{cases} 
\ee 
To perform the average over $H$ in the GOE/GUE we need the formula $\langle \exp( \frac{J}{T} {\rm Tr} (H A)) \rangle =
\exp( \frac{J^2}{\beta N T^2} {\rm Tr} A^2 )$ for $A$ symmetric/hermitian. 
Here one has $A_{ij}=\sum_{a=1}^n x_{ai} (x_{aj})^*$. Next we introduce the replica matrix and vector
\be 
Q_{ab}= \frac{1}{N} {\bf x}_a^\dagger \cdot {\bf x}_b \quad u_a = \frac{1}{N} {\bf x}_a^\dagger V {\bf x}_a
\ee 
where $Q_{ab}$ is $n \times n$ symmetric/hermitian and $u_a$ real. One has ${\rm Tr} A^2= N^2 {\rm tr} \, Q^2
= N^2 \sum_{a,b=1}^n |Q_{ab}|^2$, where here and below ${\rm tr}$ denotes the trace in the $n \times n$ replica space.
This is done via Lagrange multipliers, i.e. we introduce a $n \times n$ symmetric/hermitian
matrix $\chi_{ab}$ and $n$ real variables $w_a$ and we rewrite \eqref{ZN1} as 
\be \label{moment1} 
\! \! \langle Z_N(T)^n \rangle \propto \int du dQ dw d\chi  \, \prod_a d {\bf x}_a  \delta(Q_{aa} - 1) \, 
e^{\frac{i}{2} \sum_{ab} \chi_{ab} (N Q_{ab} - {\bf x}_a^\dagger {\bf x}_b) + 
\frac{i}{2} \sum_a w_{a} (N u_{a} - {\bf x}_a^\dagger V {\bf x}_a) + N \frac{J^2}{\beta T^2} \sum_{ab} |Q_{ab}|^2 
+ \frac{N}{T} \sum_a u_a }
\ee 
where we do not keep track of a multiplicative constant. 
Here we denote $dQ=\prod_a dQ_{aa} \prod_{a<b} dQ_{ab}$ for $\beta=1$ and
$dQ=\prod_a dQ_{aa} \prod_{a<b} d Q^r_{ab} d Q^i_{ab} $ for $\beta=2$
(the superscripts $r$ and $i$ denote real and imaginary parts),
with the same definition for $d\chi$, and $du=\prod_{a} du_{a}$, $dw=\prod_{a} dw_{a}$. 
Note that the integral over $u_a$ must be performed last. We have used for any symmetric/hermitian
matrix $B$
\be \label{deltaB} 
\delta(B_{11})= \int_{\mathbb{R}}  \frac{d\chi_{11}}{4 \pi} e^{\frac{1}{2} i \chi_{11} B_{11}}
\,  , \,  \delta(B_{12}^r) \delta(B_{12}^i)
= \int_{\mathbb{R}}  \frac{d\chi_{12}^r}{2 \pi} 
\frac{d\chi_{12}^i}{2 \pi} e^{ i \chi^r_{12} B^r_{12} - i \chi^i_{12} B^i_{12} }
= \int_{\mathbb{R}}  \frac{d\chi_{12}^r}{2 \pi} 
\frac{d\chi_{12}^i}{2 \pi}  e^{ \frac{i}{2} (\chi_{12} B_{12} + \chi_{12}^* B_{12}^*) } 
\ee 
For convergence purpose, it is important to note that the contour for the complex variable $i \chi_{11}$ 
in \eqref{deltaB} can be shifted by a real constant, as $i \chi_{11} \in 
i \mathbb{R} + \delta \chi$, and accordingly in \eqref{moment1} 
one can shift $i \chi_{ab} \to i \chi_{ab} + \delta \chi \delta_{ab}$. 
\\

We now perform the Gaussian integral over the replica fields ${\bf x}_a$ using the identity 
$\int \prod_a   d {\bf x}_a e^{- \frac{1}{2} x_{ai}^* {\cal K}_{ai,bj} x_{bj}} 
\sim e^{-\frac{\beta}{2} {\rm Tr} \log {\cal K} }$ (up to constant prefactors), where
${\cal K}_{aj,bk} = i \chi_{ab} \delta_{jk} + i V_{jk} w_a \delta_{ab}$,
where we recall that $V_{jk} = v_j \delta_{jk}$. 
We then obtain the following representation for the integer moments
\be  \label{int0} 
 \langle Z_N(T)^n \rangle \sim \int du dQ dw d\chi    \delta(Q_{aa} - 1)  e^{N S[Q,\chi,u,w] } 
\ee
in terms of the action
\be \label{SQ} 
S[Q,\chi,u,w]  =  \frac{i}{2} \sum_{ab} \chi_{ab} Q_{ab} + \sum_a 
( \frac{i}{2} w_{a} + \frac{1}{T}) u_{a} + \frac{J^2}{\beta T^2}  {\rm tr} Q^2 
- \frac{\beta}{2 N} \sum_{j=1}^N  {\rm tr} \log K_j \quad , \quad (K_j)_{ab}= i \chi_{ab} + i v_j w_a \delta_{ab}
\ee 
We have chosen the integration contours over $i \chi_{ab}$ and $i w_a$ (i.e.
$\delta \chi$ above) so that the $K_j$ are positive defined matrices. 
Until now this is exact (up to unimportant constant pre-exponential factors). 

We now assume that we can replace as $N \to +\infty$
\be 
\frac{1}{2 N} \sum_{j=1}^N  {\rm tr} \log K_j  \to  \frac{1}{2} \int dv \rho(v) {\rm tr} \log K(v) 
\quad , \quad (K(v))_{ab}= i \chi_{ab} + i v w_a \delta_{ab}
:= (i \chi + i v w)_{ab}   
\ee 
so that the action becomes 
\be \label{action} 
S[Q,\chi,u,w]  =  \frac{i}{2} \sum_{ab} \chi_{ab} Q_{ab} + \sum_a
( \frac{i}{2} w_{a} + \frac{1}{T}) u_{a} + \frac{J^2}{\beta T^2}  {\rm tr} Q^2 
- \frac{\beta}{2} \int dv \rho(v) \, {\rm tr} \log (i \chi + i v w) 
\ee 

We now assume that the integral in \eqref{int0} is dominated by a saddle point,
so that at large $N$ the asymptotics \eqref{asymptotics} holds, where $S_n(T)$ is the value of
the action $S$ at the saddle point, determined below. The saddle point (SP) equations are obtained by varying over all fields except $Q_{aa}$ which is set to unity, i.e
$Q_{aa}=1$. One obtains 
\bea \label{4SP} 
&&  \frac{\delta S}{\delta i \chi_{ab}} = 0 \,\,  \Rightarrow \quad Q_{ab} = \beta 
\int dv \rho(v) ( i \chi + i v w)^{-1}_{ab} \quad , \quad 
 \frac{\delta S}{\delta Q_{a \neq b}} =0 \, \,   \Rightarrow \quad i \chi_{a \neq b} = - \frac{4 J^2}{\beta T^2} Q^*_{a \neq b} \\
&& \frac{\delta S}{\delta i w_{a}} =0 \quad  \Rightarrow \quad u_a = \beta \int dv \rho(v) v ( i \chi + i v w)^{-1}_{aa} \quad , \quad  \frac{\delta S}{\delta u_{a}} =0 \quad  \Rightarrow \quad i w_a = - 2/T
\eea 
We note that if we enforce the last SP equation the action at the saddle point does not
depend on $u_a$, hence its value becomes immaterial in the following. However 
its value at the SP contains interesting information since it gives the overlap
of the vector with maximal eigenvalue, ${\bf x}_{\max}$,  with the matrix $V$, 
namely $u= \frac{1}{N} {\bf x}_{\max}^T V {\bf x}_{\max}$. This will be studied in
details below.

\subsection{Replica symmetric saddle point} 

We now look for a replica symmetric solution to these equations. Let us first
recall that a replica symmetric matrix can be written as
\be  
A_{ab} = A_c \delta_{ab} + A \quad , \quad A_{aa}= \tilde A=A_c+A \quad , \quad A_{a \neq b} = A 
\ee 
so we set
\bea \label{RSansatz} 
&& Q_{a \neq b} = q \quad , \quad Q_{aa}=1 \quad , \quad q_c = 1 - q \\
&& \chi_{a \neq b} = \chi \quad , \quad \chi_{aa}=\tilde \chi \quad , \quad \chi_c = \tilde \chi - \chi \nonumber \\
&& u_a=u \quad , \quad v_a = v \nonumber 
\eea 
where $q$ is real, and it turns out that at the SP $i \chi,i \tilde \chi, i  \chi_c$ are also real. 

Let us recall the formula for the inversion of RS replica matrices and their determinant
\be
(A_c \delta + A)^{-1}_{ab} = \frac{1}{A_c} \delta_{ab} - \frac{A}{A_c(A_c + n A)} \quad , \quad  \det ( A_c \delta + A) = (A_c+ n A) A_c^{n-1} 
\ee
Hence one has
\be 
( i \chi + i v w)^{-1}_{ab} = \frac{1}{i \chi_c + i v w} \delta_{ab} - \frac{i \chi}{(i \chi_c + i v w )
(i \chi_c + i v w + n i \chi)} 
\ee 
Using the last SP equation $i w=-2/T$ the first SP equation leads to 2 equations
\be \label{2eq} 
1 - q = \beta \int dv \frac{\rho(v) }{i \chi_c - 2 v/T}  \quad , \quad 
q = - \beta i \chi \int dv  \frac{\rho(v) }{(i \chi_c - 2 v/T  )
(i \chi_c + n i \chi - 2 v/T )} 
\ee 
where the first equation is obtained by taking the difference 
of \eqref{4SP} with $a=b$ and with $a \neq b$.
The second and third SP equations give
\be \label{uRS} 
i \chi = - \frac{4 J^2}{\beta T^2}  q  \quad , \quad 
u = \beta  \int dv \rho(v) v [ \frac{1}{i \chi_c - 2 v/T} - \frac{i \chi}{(i \chi_c - 2 v/T )
(i \chi_c - 2 v/T + n  i \chi)} ]
\ee 
Note that adding the two equations in \eqref{2eq} is compatible with implementing the constraint $Q_{aa}=1$ 
in the first equation of \eqref{4SP} (which thus automatically holds at the SP). 

We now eliminate $i \chi$ using the SP equation $i \chi = - \frac{4 J^2}{\beta T^2}  q$,
and for convenience we redefine
\be \label{defhatchic} 
i \chi_c =  \frac{2}{T} \kappa 
\ee 
Inserting into \eqref{2eq} we find that

(i) either $q=0$, which corresponds to a high temperature solution, in which case the real parameter $\kappa$
is determined by 
\be \label{system0} 
1 = \frac{\beta T}{2} \int dv \frac{\rho(v) }{ \kappa - v}
\ee 

(ii) or $q \neq 0$, which corresponds to the low temperature solution (the case of interest here) where $q$ and $\kappa$ are determined by the system
\bea  \label{system1} 
&& 1 = J^2   \int dv  \frac{\rho(v) }{(\kappa - v  )
(\kappa - v - 2 n J^2 q/(\beta T))} \\
&& 1 - q = \frac{\beta T}{2} \int dv \frac{\rho(v) }{\kappa - v}  \label{system2} 
\eea 
\\

On the other hand, using the ansatz \eqref{RSansatz}, one can rewrite the action \eqref{action} in the RS subspace as 
$S[Q,\chi,u,w]|_{RS}=S[q,\chi,\chi_c]$ with
\bea 
 \frac{1}{n} S[q,\chi,\chi_c]   &=& \frac{1}{2} ( i \chi_c + i \chi (1+  (n-1) q)  )
 + \frac{J^2}{\beta T^2}   (1 + (n-1) q^2) \\
& -& \frac{\beta}{2 n} \int dv \rho(v) \left( (n-1) \log(i \chi_c - 2 v/T) + \log(i \chi_c - 2 v/T + n i \chi) \right) \nn
\eea 
where we have not yet inserted the values of the parameters $q, i \chi_c, i \chi$ at the SP. 
For consistency, one can check (see Appendix \ref{app:derivation}) that if one takes the derivatives of the RS action
$S[q,\chi,\chi_c]$ w.r.t. the parameters $q,\chi,\chi_c$ one recovers the RS version
of the general SP equations given above. 

Let us now use one of the saddle point equations by inserting $i \chi = - \frac{4 J^2}{\beta T^2}  q$
and use the variable $\kappa$ defined in \eqref{defhatchic}. One then obtains the action 
$S[q,\chi,\chi_c]=S[q,\kappa]+\frac{\beta n}{2} \log(T/2)$ as a function of $\kappa$ and $q$ as
\be \label{actionT} 
 \frac{1}{n}  S[q,\kappa]  =   \frac{\kappa}{T} +  \frac{J^2}{\beta T^2}   (1 -2 q - (n-1) q^2)  
- \frac{\beta}{2 n} \int dv \rho(v) \left( (n-1) \log(\kappa - v) + \log(\kappa - v -  2 n J^2 q/(\beta T) ) \right) 
\ee 
Note that if we now take derivatives w.r.t. $q$ and $\kappa$ 
\bea \label{twoeq} 
\!\!\!\! \!\! \!\!  && \frac{\delta S[q,\kappa]}{\delta q} =0 \quad \Leftrightarrow \quad 1 + (n-1) q = \frac{\beta T}{2} \int dv \rho(v)  
\frac{1}{\kappa - v -   2 n J^2 q/(\beta T)} \label{der1} \\
\!\!\!\! \!\! \!\!   && \frac{\delta S[q,\kappa] }{\delta \kappa} \quad \Leftrightarrow \quad 1 -  \frac{\beta T}{2} \int dv \rho(v) \frac{1}{\kappa -  v} 
+ \frac{\beta T}{2 n}  \int dv \rho(v) \left( \frac{1}{\kappa - v } - \frac{1}{\kappa - v -  2 n J^2 q/(\beta T)} 
\right) = 0 
\eea 
One can check that these two equations, for $q \neq 0$, are two independent linear combinations of the SP equations 
\eqref{system1}-\eqref{system2}, and for $q=0$, are equivalent to \eqref{system0}. 
Inserting their solution in the action gives $S_n(T)=S[q,\kappa]|_{SP}$, see below. 
The representation \eqref{actionT} of the action is convenient for the following.
 
Note that one must have $\kappa \geq v_e$, where $v_e$ is the upper edge of $\rho(v)$, for considerations of positivity of the matrix $K$ in \eqref{SQ}. 
\\

 \noindent
{\bf Absence of deformation $V=0$}. Let us consider the case of pure GOE/GUE ensemble $V=0$, i.e. $\rho(v)=\delta(v)$. 
The system \eqref{system1}-\eqref{system2} leads to (for $q \neq 0$)
\be 
1 - q = \frac{\beta T}{2 \kappa}  \quad , \quad 1 = \frac{J^2}{\kappa (\kappa  - 2 n J^2 q/(\beta T))} 
\ee 
Eliminating $\kappa$ using the first equation one finds the equation which determines $q$
\be  
J^2 (1-q) (1 + (n-1) q) = \frac{\beta^2 T^2}{4}
\ee  
which coincides with Eq. (29) in \cite{TrivializationUs2014}, together with the additional solution $q=0$. 
Because of different conventions, to compare we must change our $T \to 2 T$ and set $\beta=1$
for GOE. For $n=0$ one recovers (in our conventions) the spin glass order parameter as
\be 
q= q_0(T) = \max( 0 , 1- \frac{T}{T_c} ) \quad , \quad T_c = \frac{2 J}{\beta}
\ee 
where $T_c$ is the transition temperature of the spherical spin glass. 
\\

\subsection{Typical value of $\lambda_{\rm max}(M)$}

We first study the limit $n=0$ which gives the mean free energy of the model at any temperature,
as well as the typical value 
$\lambda_{\rm typ}$ of the largest eigenvalue $\lambda_{\rm max}(M)$ of $M$ in the limit $T=0$.
\\

Let us set $n=0$ in the SP equations \eqref{system1}-\eqref{system2} 
which are valid in the low temperature phase $T<T_c$. One obtains
\be \label{sptyp1} 
1 - q = \frac{\beta T}{2}  \int dv \frac{\rho(v) }{\kappa - v}  \quad , \quad 
1 = J^2 \int dv  \frac{\rho(v) }{(\kappa - v  )^2} 
\ee 
Then two cases can occur.
\\

{\bf First case}. Consider first the case I where $\rho(v)$ is such that
\be 
 \int dv  \frac{\rho(v) }{(v_e - v  )^2} = + \infty
\ee 
Then the second equation in \eqref{sptyp1} has always a unique root with $\kappa > v_e$.
This root is independent of $T$ and one has
\be  \label{59} 
q = 1 - T \hat q \quad , \quad \hat q = \frac{\beta}{2} \int dv \frac{\rho(v) }{\kappa - v} 
\ee 

Let us now compute the mean free energy 
\be 
\langle {\sf f}(T) \rangle =  \lim_{N \to +\infty} \frac{T}{N} \langle \log Z_N(T) \rangle = \lim_{N \to +\infty, n \to 0} 
 \frac{T}{N n} \log \langle Z_N(T)^n \rangle = T \lim_{n \to 0} \frac{1}{n} S_n(T)
\ee 
where here and below $S_n(T)=S[q,\kappa]$ denotes the action at the saddle-point. 
Taking the limit $n \to 0$ of \eqref{actionT} one finds 
\be \label{actionT2} 
\langle {\sf f}(T) \rangle = T \lim_{n \to 0} \frac{1}{n} S_n(T)  =   \kappa +  \frac{J^2}{\beta T}  (1-q)^2 
- \frac{\beta T }{2} \int dv \rho(v) 
\left( \log(\kappa - v) - \frac{2 J^2 q/(\beta T)}{\kappa - v} \right)
\ee 
where $q$ and $\kappa$ are given by the solution of the system \eqref{sptyp1}.

We note that, leaving aside the possibility of first order transitions, 
the transition from low to high temperature (if it is continuous) now occurs at a value of $T=T_c$
obtained by eliminating $\kappa$ in the system 
\be \label{sptyp1n} 
1  = \frac{\beta T_c}{2}  \int dv \frac{\rho(v) }{\kappa - v}  \quad , \quad 
1 = J^2 \int dv  \frac{\rho(v) }{(\kappa - v  )^2} 
\ee

From the mean free energy \eqref{actionT2}, by taking the limit $T \to 0$ we obtain the mean, i.e. the typical, largest eigenvalue of $M$ as
as 
\be
\langle \lambda_{\max}(M)  \rangle = \lim_{T \to 0}  \langle {\sf f}(T) \rangle = 
\kappa 
+  J^2  \int dv  \frac{\rho(v)}{\kappa - v} 
\ee
where we used that $1-q=O(T)$ from \eqref{59}. In this formula $\kappa$ is the
root of the second equation in \eqref{sptyp1}.
\\

In the absence of deformation $V=0$, i.e. for the GOE/GUE one finds that 
\be  
\kappa=J \quad , \quad q = 1 - \frac{\beta T}{2 J}  \quad , \quad \langle \lambda_{\max}(M)  \rangle = 2 J
\ee   
which reproduces the correct right edge of the semi-circle spectrum.
\\

To summarize, we find that in this first case, the typical value $\lambda_{\rm typ}$ of $\lambda_{\rm max}(M)$ 
is given by eliminating $\kappa$ in the system
\bea \label{summ1} 
&& \lambda^{\rm typ}_{\max}(M) = \langle \lambda_{\max}(M)  \rangle = \kappa + J^2  \int dv  \frac{\rho(v)}{\kappa - v}  \\
&& 1 = J^2 \int dv  \frac{\rho(v) }{(\kappa - v  )^2} 
\eea 
This is equivalent to the formulation given in \eqref{sec:main}, see e.g. \eqref{typical},
with $z^*=\kappa$, valid in the delocalized phase. 

Note that in \cite{Krajenbrink2021} not only the mean value but the fluctuations were also studied.
One sees that Eqs. \eqref{summ1} are identical with Eqs (80-82) there, with the correspondence in notations
$\mu= \lambda_{\max}( \sqrt{\theta} H + \theta V) = \theta \lambda_{\max}(M)|_{J=1/\sqrt{\theta}}$
and $z^* = \kappa$. 
\\

{\bf Second case}. Consider now the case II where $\rho(v)$ is such that
\be  \label{cas2} 
 \int dv  \frac{\rho(v) }{(v_e - v  )^2} < +\infty
\ee 
Then, as $J$ is decreased from $+\infty$, we see that the root $\kappa$ of  \eqref{sptyp1}
decreases until it reaches $\kappa = v_e$ at $J=J_c$ given by
\be 
J_c^2 = \frac{1}{\int dv  \frac{\rho(v) }{(v_e - v  )^2}} 
\ee 
For $J>J_c$ all the results of the previous case I still apply. In terms of the
directed polymer this corresponds to the {\it delocalized phase}. The case I above corresponds
$J_c=0$, so that there is only a delocalized phase.
\\

For $J < J_c$, that is in the {\it localized phase} for the polymer, the value of $\kappa$ at the saddle point freezes at $\kappa = v_e$. Indeed it cannot
be smaller. A more detailed analysis, as the one performed in \cite{Krajenbrink2021}, shows that it approaches $v_e$ up to
terms of order $1/N^a$, but here we are not studying such refined scales (it require to
go beyond the simplest SP analysis). 

Since the value of $\kappa$ is frozen, we cannot use anymore that the variation of the RS action w.r.t. $\kappa$ vanishes. It implies that the second equation in \eqref{sptyp1} is not valid anymore.
Hence we are left with the single equation \eqref{der1}, and inserting there for $n=0$
the value $\kappa=v_e$ determines
the value of $q$ as
\be \label{sptyp2} 
q = 1- \frac{\beta T}{2}  \int dv \frac{\rho(v) }{v_e - v} 
\ee 
where the integral is finite since one has \eqref{cas2}. Ignoring again possible first order transitions, the low to high temperature transition is now occuring, for $J<J_c$, at
\be 
T_c = \frac{2}{\beta} \frac{1}{\int dv  \frac{\rho(v) }{v_e - v  }} 
\ee 
i.e. it is independent of $J$. The mean free energy in the low temperature phase is now given by
\be \label{actionT3} 
\langle {\sf f}(T) \rangle = T \lim_{n \to 0} \frac{1}{n} S_n(T)  =   v_e +  \frac{J^2}{\beta T}  (1-q)^2 
- \frac{\beta T }{2} \int dv \rho(v) 
\left( \log(v_e - v) - \frac{2 J^2 q/(\beta T)}{v_e - v} \right)
\ee 
where $q$ is given by \eqref{sptyp2}. Taking now the limit $T \to 0$ one 
obtains the mean, i.e. the typical, largest eigenvalue of $M$ in the localized
phase $J<J_c$ as
\be
\langle \lambda_{\max}(M)  \rangle = \lim_{T \to 0}  \langle {\sf f}(T) \rangle = \lambda^{\rm typ,loc} =
v_e 
+  J^2  \int dv  \frac{\rho(v)}{v_e - v} 
\ee


\subsection{Large deviation scaled CGF $\phi(s)$}

We now obtain the large deviation scaled CGF $\phi(s)$ defined in \eqref{defphi1}, which according to \eqref{defphi2}
can be obtained from the action at the saddle point $S_n(T)$ as 
\be  \label{PhiLim} 
\phi(s) = \lim_{T \to 0} S_{n=s T}(T)
\ee
We will restrict to $s \geq 0$. 

Thus from now on we set $n = T s$ and take simultaneously the $n=0$ and $T=0$ limit at fixed $s$. 
The saddle point equation \eqref{system2} shows that $q \to 1$, with $1-q=O(T)$ more precisely
\be 
1-q = T \hat q \quad , \quad \hat q = \frac{\beta}{2} \int dv \frac{\rho(v) }{\kappa - v} 
\ee 
Inserting $q \to 1$ in  \eqref{system1} leads to the equation
\be \label{cond1} 
1 = J^2 \int dv  \frac{\rho(v) }{(\kappa -  v  )
(\kappa - v - 2 s J^2/\beta)} 
\ee 
which determines $\kappa= \kappa(s)$ as a function of $s$. 
Similarly to the previous section, \eqref{cond1} is valid only if its root $\kappa= \kappa(s)$ satisfies both
\be \label{2cond} 
\kappa \geq v_e  \quad , \quad \kappa - 2 s J^2/\beta \geq v_e 
\ee 
We will call this the {\it delocalized regime} of the large deviations, which defines some domain of
the $(J,s)$ plane determined below. For $s=0$ it is equivalent to
the delocalized phase of the model $J \geq J_c$. It turns out that the delocalized regime
is a subdomain of the delocalized phase. So we will focus now on $J \geq J_c$.

Let us determine the boundaries of this delocalized regime. For $J \geq J_c$, that is in the delocalized phase,
let us define $s^*=s^*(J) \geq 0$ the unique root of
\be \label{sstar} 
1 = J^2 \int dv  \frac{\rho(v) }{(v_e + 2 s^* J^2/\beta -  v  )(v_e - v)}  
\ee 
where we assume here that $\int dv  \frac{\rho(v) }{v_e - v} < +\infty$. 
For $J \to J_c^+$ one has $s^* \to 0^+$ and for $J \to +\infty$ one sees
that $s^* \to \int dv  \frac{\rho(v) }{v_e - v}$. Then we see that for 
$0 \leq s \leq s^*$ Eq. \eqref{cond1} has a unique root $\kappa=\kappa(s)$ in the interval 
$\kappa \in [v_e + 2 s J^2/\beta, + \infty[$. This is the delocalized regime. 
As $s$ reaches $s^*(J)$ one sees that $\kappa(s)$ reaches 
$v_e + 2 s J^2/\beta$ and one exits from the delocalized regime. 

Within the delocalized regime, $J \geq J_c$ and $0 \leq s \leq s^*$, we 
obtain, from \eqref{actionT} and \eqref{PhiLim}, the scaled CGF as
\be \label{actionT0} 
 \phi(s) =  \phi(s, \kappa(s)) \quad , \quad 
 \phi(s, \kappa) :=
  \kappa s - \frac{J^2 s^2}{\beta}  
- \frac{\beta}{2} \int dv \rho(v)  \log\left( \frac{\kappa - v - 2 s J^2/\beta}{\kappa - v} \right)  
\ee
where $\kappa=\kappa(s)$ is the solution of \eqref{cond1}
assumed to satisfy \eqref{2cond}. Note that Eq. \eqref{cond1} can be rewritten as
\be \label{dercond} 
\partial_{\kappa} \phi(s, \kappa)|_{\kappa=\kappa(s)} = 0
\ee 

Let us consider now, still within the delocalized phase, the region $s \geq s^*$.
This is {\it the localized regime} of large deviations within the delocalized phase. Similarly to the previous section, $\kappa$ 
freezes at its boundary value (which is now $s$ dependent)
\be \label{frozenchi} 
 \kappa = v_e + 2 s J^2/\beta
\ee
Then \eqref{cond1} does not hold anymore and one can use only the SP equation
\eqref{der1} which reads, upon inserting $s= n T$ and \eqref{frozenchi}, to
lowest order as $T \to 0$,
\be
1 - q = T \left( -  s + \frac{\beta T}{2} \int dv \rho(v)  
\frac{1}{v_e - v } \right) + O(T^2) 
\ee  
It turns out that below we do not need the precise value of the $O(T)$ term. 

We can now inject this value of $q$ together with $\kappa$ from \eqref{frozenchi} 
into the expression for the action \eqref{actionT} and obtain using 
\eqref{PhiLim} the scaled CGF $\phi(s)$ for $J \geq J_c$ and $s \geq s^*$, i.e. in the localized regime
of the delocalized phase as
\be  \label{Philocreg} 
\phi(s)  = \phi(s,v_e + 2 s J^2/\beta) = 
  v_e s + \frac{J^2 s^2}{\beta}  
- \frac{\beta}{2} \int dv \rho(v)  \log\left( \frac{v_e - v }{v_e - v + 2 s J^2/\beta} \right)  
\ee 
where $\phi(s,\kappa)$ is given by the same formula as 
\eqref{actionT0}.
\\

In the localized phase, i.e for $J \leq J_c$, it is formally as if $s^*=0$, so there is only the
localized regime and $\phi(s)$ is given 
again by the expression \eqref{Philocreg} for all $s \geq 0$.
\\

The convexity of $\phi(s)$ is shown in Appendix \ref{app:convex}. 

%

 \noindent
{\bf Absence of deformation $V=0$}. In the case of pure GOE/GUE $\rho(v)=\delta(v)$
there is only a delocalized phase and a delocalized regime. One finds 
that the only solution of \eqref{cond1} which obeys $\kappa(s) > 2 s J^2/\beta$ is 
given by
\be 
\kappa(s) =  \frac{J^2}{\beta} (s +  \sqrt{s^2 + \frac{\beta^2}{J^2} } )
\ee 
which leads to
\be  
\phi(s) = \frac{J^2}{\beta} s  \sqrt{s^2 + \frac{\beta^2}{J^2} } + \beta \log(\frac{s + \sqrt{s^2 + \frac{\beta^2}{J^2} }}{\beta/J}) 
\ee   
which coincides with the result (40) in \cite{TrivializationUs2014} (upon setting $s \to 2 s$ there). 
\\

\section{Rate function ${\cal L}(\lambda)$}

\subsection{General formula} 

Having obtained $\phi(s)$ we can now obtain the rate function 
which governs the upper tail of the PDF $P(\lambda)$ of $\lambda= \lambda_{\rm max}(M)$.
Assuming the large deviation form on the side $\lambda \geq \lambda_{\rm typ}$ 
\be \label{LD1n} 
P(\lambda) \sim e^{- \beta N {\cal L}(\lambda) } 
\ee 
Then one has, for $s \geq 0$
\be
\langle e^{N s {\lambda}_{\max}(M) } \rangle = \int d\lambda P(\lambda) e^{N s \lambda }
\sim \int_{\lambda \geq \lambda_{\rm typ}} d\lambda e^{ N (s \lambda - \beta {\cal L}(\lambda)) } 
\ee 
since for $s \geq 0$ we expect the integral to be dominated by a saddle point within the upper tail 
$\lambda \geq \lambda_{\rm typ}$. Comparing with \eqref{defphi1} we deduce
\be 
\phi(s) = \max_{\lambda \geq \lambda_{\rm typ}}  (s \lambda - \beta {\cal L}(\lambda)) 
\ee 
The rate function ${\cal L}(\lambda)$ for $\lambda \geq \lambda_{\rm typ}$ 
can then be obtained from the Legendre transform 
\be \label{max} 
\beta {\cal L}(\lambda) = \max_{s \geq 0} ( s \lambda -  \phi(s)) 
\ee  

To construct ${\cal L}(\lambda)$, and assuming that the extremum is not on a boundary,
we write the derivative conditions
\be  \label{dercond2} 
s = \beta {\cal L}'(\lambda)   \quad , \quad \lambda = \phi'(s) 
\ee 
which relate $s$ and $\lambda$.
We consider the two cases. 
\\

{\bf Rate function in the delocalized regime}. Consider the delocalized phase $J>J_c$, i.e. \eqref{sstar} has a strictly positive root 
$s^*>0$. Within the delocalized phase, consider the delocalized regime $0 \leq s \leq s^*$. From \eqref{actionT0} 
and using \eqref{dercond} the derivative condition gives
\be \label{e1} 
\lambda =  \phi'(s)  = \partial_s \phi(s, \kappa)|_{\kappa=\kappa(s)} 
=  \kappa - \frac{2 J^2}{\beta} s
+ J^2 \int dv  \frac{\rho(v)} {\kappa - v - 2 s J^2/\beta} 
\ee 
where $\kappa=\kappa(s)$ is the solution of \eqref{cond1}
which in the delocalized regime satisfies \eqref{2cond}. 
Using \eqref{cond1} it can be simplified into
\be \label{simpler} 
\lambda
=  \kappa(s) 
+  J^2  \int dv  \frac{\rho(v)} {\kappa(s) - v} 
\ee
which, remarkably, is the same relation as \eqref{e1} upon performing the substitution $\kappa \to \kappa - 2 s J^2/\beta$ in \eqref{simpler}. 
It is thus convenient to define an auxiliary function $f(z)$ for $z \in [v_e,+\infty[$,
and to consider the equation
\be \label{eqeq} 
\lambda = f(z) := z
+ J^2  \int dv  \frac{\rho(v)} {z - v} 
\ee 
The function $f(z)$ is convex. It diverges for $z \to +\infty$ and 
attains for $z=v_e$ a finite value $f(v_e)=\lambda_e$. 
In the delocalized phase $J > J_c$, the unique minimum
of $f(z)$ is at $z=z^* > v_e$ with $f(z^*)=\lambda_{\rm typ}$. For 
$\lambda_{\rm typ} < \lambda < f(v_e)$ the equation \eqref{eqeq} 
admits two roots $z_+=z_+(\lambda)= \kappa$ and $z_-=z_-(\lambda) = \kappa - 2 J^2 s/\beta \leq z_+$
(they become equal for $\lambda = \lambda_{\rm typ}$ where $z_\pm(\lambda_{\rm typ})=z^*$). 
This corresponds to the delocalized regime, with $0< s<s^*$.
Within this regime, using \eqref{max}, inserting $\lambda=f(\kappa)$ according to \eqref{simpler} 
and using \eqref{actionT0}, one finds that the rate function is
\be \label{Ldeloc} 
\beta {\cal L}(\lambda) = s \lambda -  \phi(s) 
=  \frac{J^2}{\beta} s^2 + s J^2 \int dv   \frac{\rho(v)}{\kappa(s) - v}  
+\frac{\beta}{2} \int dv  \rho(v) 
    \log\left( 1 -  \frac{2 s J^2/\beta}{\kappa(s) - v} \right)  
\ee 
where $\kappa=\kappa(s)$ is the solution of \eqref{cond1}
which satisfies \eqref{2cond}, and $s$ is related to $\lambda$ through \eqref{e1}.
Using that
\be \label{101}
2 J^2 s/\beta = z_+ - z_- \quad , \quad  \kappa= z_+ \quad , \quad \kappa - 2 J^2 s/\beta = z_-
\ee 
The rate function in the delocalized regime, $J > J_c$ and $0< s<s^*$, can be rewritten as
\bea 
  {\cal L}(\lambda)  
&=&  \frac{1}{4 J^2} (z_+ - z_-)^2 +  \frac{1}{2} \int dv  \rho(v) \bigg[ \frac{z_+ - z_-}{z_+ - v} 
+    \log \frac{z_- -v}{z_+ - v}   \bigg] \\
&=&  \frac{1}{4 J^2} ( z_+-z_-) (2 \lambda - z_+ - z_-)  
+  \frac{1}{2} \int dv  \rho(v) \log \frac{z_- -v}{z_+ - v} \label{ratedeloc1} 
\eea 
where $z_\pm=z_\pm(\lambda)$ are the two roots of \eqref{eqeq}. 
\\

{\bf Rate function in the localized regime}. Let us first consider the delocalized phase $J > J_c$
where $s^*>0$. 
From the above discussion, the smallest root $z_-(\lambda)$ of \eqref{eqeq} reaches $z_-=v_e$
when $\lambda = f(v_e) = \lambda_e$. This corresponds to
$s$ reaching $s^*$ (by comparing \eqref{frozenchi} and \eqref{101}, or see below).
For $\lambda >  f(v_e)$ the rate function is
determined by the localized regime $s > s^*$ of $\phi(s)$. 
In that regime the root $z_-$ has disappeared, and there remains only 
the single root $z_+$ to the equation $\lambda=f(z)$. This root 
furthermore obeys $z_+(\lambda) = \kappa(s) = v_e + 2 s J^2/\beta$.
Indeed, in that regime the scaled CGF $\phi(s)  = \phi(s,v_e + 2 s J^2/\beta)$ is given by
\eqref{Philocreg}, and taking a derivative to perform the Legendre transform
we obtain 
\be  \label{104} 
\lambda= \phi'(s) = v_e + \frac{2 J^2 s}{\beta} + J^2 \int dv  \frac{\rho(v)} {v_e - v + 2 s J^2/\beta} 
= f(v_e + \frac{2 J^2 s}{\beta}) 
\ee 
Note that at $s \to s^*$ using \eqref{sstar} and comparing with \eqref{eqeq} this equation gives,
as expected, 
$\lambda = \lambda_e = f(v_e)=x_c$.
One then obtains the rate function in the delocalized regime $s > s^*$ as
\be
\beta {\cal L}(\lambda) = s \lambda -  \phi(s) 
=   \frac{J^2 s^2}{\beta} 
+ J^2 s \int dv  \frac{\rho(v)} {v_e - v + 2 s J^2/\beta} 
+ \frac{\beta}{2} \int dv \rho(v)  \log\left( \frac{v_e - v}{v_e - v+ 2 s J^2/\beta} \right)  
\ee
where $s$ is related to $\lambda$ via Eq. \eqref{104}.
The continuity of this expression with \eqref{Ldeloc} at $s=s^*$ can be seen
using that $\kappa=v_e + 2 s J^2/\beta$ at $s=s^*$. Using that 
$s=\frac{\beta}{2 J^2}(z_+-v_e)$ and \eqref{104} the result
can be written in the equivalent form
\be \label{rateloc2} 
  {\cal L}(\lambda)  =  \frac{1}{4 J^2}  (z_+-v_e) (2 \lambda - z_+ - v_e) 
+ \frac{1 }{2} \int dv \rho(v)  \log\left( \frac{v_e - v}{z_+ - v} \right)  
\ee
which is formally identical to the expression \eqref{ratedeloc1} for the rate function in delocalized
phase, upon freezing $z_- = v_e$. 

Finally in the localized phase $J<J_c$, $\phi(s)$ is always in the localized
regime. The function $f(z)$ reaches its minimum in the interval $[v_e,+\infty[$ at 
its lower edge $z=v_e$. The typical value of $\lambda$ is then 
$\lambda_{\rm typ, \rm loc}=f(v_e)$. For $\lambda > f(v_e)$ the 
relation between $s$ and $\lambda$ is still given by \eqref{104} 
and the 
rate 
function has the same expression \eqref{rateloc2}. 
\\

\noindent
{\bf Case $V=0$}. As a check, let us consider GOE/GUE matrices without perturbation. One has
\be  
f(z) = z + \frac{J^2}{z} \quad , \quad z^*=J \quad , \quad z_\pm(\lambda)= \frac{1}{2} (\lambda \pm \sqrt{\lambda^2 - 4 J^2}) 
\ee  
Thus for $\lambda \geq \lambda_{\rm typ}=2 J$
\be  \label{standard} 
{\cal L}'(\lambda) = \frac{1}{2 J^2}  \sqrt{\lambda^2 - 4 J^2} \quad , \quad 
{\cal L}(\lambda) = \frac{\lambda}{4 J^2}  \sqrt{\lambda^2 - 4 J^2} - \log( \frac{\lambda + \sqrt{\lambda^2 - 4 J^2}}{2 J}) 
\ee 
which is the standard result \cite{BenArous2001,MajumdarVergassola},
obtained by the present method in \cite{TrivializationUs2014}. Near the typical value, for 
$\frac{\lambda}{J}-2=O(1)$, it behaves as
\be \label{expTW} 
{\cal L}(\lambda) \simeq \frac{2}{3} (\frac{\lambda}{J}-2)^{3/2} + \frac{1}{20} (\frac{\lambda}{J}-2)^{5/2}
\ee 
The $3/2$ exponent there signals the "third order" phase transition at the upper edge of the spectrum of the GOE/GUE random
matrices \cite{SatyaGregRMTReview}. As discussed in \cite{TrivializationUs2014}, the form \eqref{expTW} 
matches (including the coefficient) the upper tail of the GOE/GUE Tracy-Widom distributions
$F'_\beta(\chi) \sim e^{- \beta \frac{2}{3} \chi^{3/2}}$, where $\chi= N^{2/3}( \frac{\lambda}{J}-2)=O(1)$, which
describe the typical fluctuations of order $O(N^{-2/3})$ of $\lambda_{\rm max}(M)$ (and cannot be accessed
by the present method). We generalize this result to the deformed GOE/GUE below in \eqref{serexp}. 
\\

\subsection{Behavior near the typical value} \label{subsec:behavior} 
We now use the general formula for the rate function to study its behavior near the typical value.
In the delocalized phase $J>J_c$ it is useful to rewrite \eqref{ratedeloc1} as ${\cal L}(\lambda) = {\cal L}(\lambda,z_+(\lambda),z_-(\lambda))$. One 
finds 
\bea \label{derzpm} 
& \partial_{z_\pm} {\cal L}(\lambda,z_+,z_-) =  \pm \frac{1}{2 J^2} (\lambda - f(z_\pm)) 
\eea 
Since by definition $f(z_\pm)=\lambda$ one has, as expected (since both sides are equal to $s/\beta$)
\be \label{Lprime} 
{\cal L}'(\lambda) = \frac{1}{2 J^2} (z_+(\lambda)-z_-(\lambda) )
\ee 

Let us recall that the typical value is $\lambda_{\rm typ}=f(z^*)$, where $z^*$ is the unique root of $f'(z^*)=0$ for $z^* > v_e$.
To obtain the behavior of the rate function near the typical value, rather than using
 \eqref{ratedeloc1}, it is easier to 
expand $z_\pm(\lambda)$ in powers of $\lambda-\lambda_{\rm typ}$ and integrate \eqref{Lprime}.
One uses the expansion $\lambda-\lambda_{\rm typ}= \frac{1}{2} f''(z^*)  (z_\pm(\lambda)-z^*)^2 + \dots$
and obtains
\be \label{serexp} 
{\cal L}(\lambda) = 
\frac{2 \sqrt{2} }{3 J^2
   \sqrt{f''(z^*)}} (\lambda - \lambda_{\rm typ})^{3/2}
     +\frac{5
   f^{(3)}(z^*)^2-3 f^{(4)}(z^*)
   f''(z^*)}{45 \sqrt{2} J^2
   f''(z^*)^{7/2}} (\lambda - \lambda_{\rm typ})^{5/2}
   + O( (\lambda - \lambda_{\rm typ})^{7/2} )
   \ee 
The above expansion assumes that the derivatives $f^{(p)}(z^*)$ for
$p \geq 2$ exist, which is true in the delocalized phase (for a smooth density), since $z^*>v_e$.
 In view of the discussion following Eq. \eqref{expTW}, it
 is reasonable here to find again the $3/2$ exponent. Indeed, it was found
 in \cite{Krajenbrink2021} in the GUE case that the typical fluctuations in the
 delocalized phase are still described by the Tracy-Widom distribution. 
 The present results shows that for $\lambda - \lambda_{\rm typ} = O(1)$ one has
 \be  \label{tailTW} 
 P(\lambda) \sim e^{- \beta N \frac{2 \sqrt{2} }{3 J^2
   \sqrt{f''(z^*)}} (\lambda - \lambda_{\rm typ})^{3/2} } 
 \ee 
Matching amounts
to assume the absence of any intermediate regime, hence that \eqref{tailTW} also describes
the upper tail of the typical fluctuations 
$\lambda - \lambda_{\rm typ} = O(N^{-2/3})$. 
One can check that it again works, including the coefficient, if compared 
to Eq. (113) in \cite{Krajenbrink2021} (where $\varphi^{(3)}(z^*)$
there equals $f''(z^*)/J^2$). It was shown there that
$\chi= (2 J^2/f''(z^*))^{1/3} N^{2/3} (\lambda - \lambda_{\rm typ})/J^2=O(1)$
follows the Tracy Widom distribution $F_\beta$ for $\beta=2$. 
\\

In the localized phase $J<J_c$, one writes ${\cal L}(\lambda) = {\cal L}(\lambda,z_+(\lambda),v_e)$,
and since \eqref{derzpm} is still valid for $z_+$, one finds that \eqref{Lprime} is replaced by
\be \label{Lprime2} 
{\cal L}'(\lambda) = \frac{1}{2 J^2} (z_+(\lambda)-v_e )
\ee 
The typical value is now $\lambda_{\rm typ}=\lambda_e=f(v_e)$, and one can expand
$\lambda-\lambda_e=f'(v_e) (z_+-v_e) + \dots$. One finds
\be \label{Lloc} 
{\cal L}(\lambda) = \frac{(\lambda-\lambda_e)^2}{4 J^2
   f'\left(v_e\right)}-\frac{ f''\left(v_e\right)}{12 J^2 f'\left(v_e\right){}^3} (\lambda-\lambda_e)^3
   +o\left((\lambda-\lambda_e)^3\right)  \quad , \quad f'(v_e)=1 - \frac{J^2}{J_c^2} >0
\ee  
Hence in the localized phase the large deviation rate function shows a Gaussian behaviour around the
typical value. Extrapolating from the large deviation tail one infers
\be \label{varianceloc} 
{\rm Var} (\lambda-\lambda_e) = \frac{2 J^2}{\beta N} (1 - \frac{J^2}{J_c^2})
\ee 
hence that the fluctuations of $\lambda=\lambda_{\rm max}(M)$ should be $O(1/N^{1/2})$. 
We can compare with the upper tail of the typical fluctuations which were studied in 
Ref. \cite{Krajenbrink2021}, and found indeed to be Gaussian, see Eq. (107-109) there, with the correspondence 
$\lambda_1/N=\lambda/J^2$, $\theta=1/J^2$ and one must use $\beta=2$. 
The matching works, but surprisingly only up to a factor of $2$. Indeed, we
find that there would be a perfect correspondence with \eqref{varianceloc} if the variable
$\omega$ there had a variance ${\rm Var}(\omega)$ equal to $1/2$, while
Eq. (109) there implies that ${\rm Var}(\omega)=1/4$. The origin of this
factor remains to be understood. 
\\

Note that in \eqref{Lloc} we have assumed $f''(v_e) = 2 \int dv \frac{\rho(v)}{(v_e-v)^3} < +\infty$. 
For densities which vanish at the upper edge as $\rho(v) \sim (v_e-v)^\alpha$ it means $\alpha>2$. 
If $1 < \alpha <2$ the second term in \eqref{Lloc} will be $\sim (\lambda-\lambda_e)^{1+\alpha}$. 
We assume here that $\alpha>1$, i.e. $f'(v_e)$ is finite which implies that $J_c>0$. 
If $\alpha<1$ there is no localized phase.

{\bf Interpolation around the critical point}. As $J$ is decreased close to $J=J_c$ the rate function ${\cal L}(\lambda)$
should interpolate between its form in the delocalized phase $J>J_c$, i.e. Eq. \eqref{serexp}, and its
form in the localized phase $J<J_c$, i.e. \eqref{Lloc}. This requires a separate study, performed in the
Appendix \ref{app:interpolation}. Let us display here only the result. 

The critical region is defined by $|f'(v_e)| \ll 1$, where $f'(v_e)=1 - \frac{J^2}{J_c^2} \simeq \frac{2 (J_c-J)}{J_c}$,
and $\lambda-\lambda_{\rm typ} = O((J-J_c)^2)$. 

On the delocalized side, $J>J_c$, $f'(v_e)<0$, and within the critical region, there are two regimes
separated by $\lambda=\lambda_e=f(v_e)$ with
\be
\lambda_e-\lambda_{\rm typ} \simeq  \frac{f'(v_e)^2}{2 f''(v_e)} = O(|J_c-J|^2) 
\ee 
where the rate function takes the respective forms
\bea
&& {\cal L}(\lambda) \simeq \frac{2}{3 J^2} \sqrt{\frac{2}{f''(v_e)}} (\lambda - \lambda_{\rm typ})^{3/2}  \quad , \quad 
\lambda_{\rm typ} < \lambda < \lambda_e \\
&& {\cal L}(\lambda) \simeq \frac{1}{3 J^2} \sqrt{\frac{2}{f''(v_e)}} (\lambda - \lambda_{\rm typ})^{3/2} 
+ \frac{|f'(v_e)|}{2 J^2 f''(v_e)} (\lambda - \lambda_{\rm typ})  - \frac{1}{12 J^2} \frac{|f'(v_e)|^3}{f''(v_e)^2} 
 \quad , \quad 
\lambda > \lambda_e
\eea
where $J$ can be replaced by $J_c$ and $J^2 f''(v_e)$ and $\lambda/J$ are dimensionless.

On the localized side, $J<J_c$, $f'(v_e)>0$, one has $\lambda_{\rm typ}=\lambda_e=f(v_e)$ and the rate function
takes the form
\bea
&& {\cal L}(\lambda) \simeq \frac{1}{3 J^2} \sqrt{\frac{2}{f''(v_e)}} (\lambda - \lambda_{e} + \frac{f'(v_e)^2}{2 f''(v_e)} )^{3/2} 
- \frac{f'(v_e)}{2 J^2 f''(v_e)} (\lambda - \lambda_{e})  - \frac{1}{6 J^2} \frac{|f'(v_e)|^3}{f''(v_e)^2} 
\eea

\section{Overlap and occupation measure}

Let us denote ${\bf x}_{\max}$ the eigenvector associated to 
the largest eigenvalue $\lambda_{\rm max}(M)$  of $M$. It is
normalized as ${\bf x}_{\max}^T {\bf x}_{\max}=N$. Consider the
overlap of this vector with the perturbation $V$, which we define as
\be 
u := \frac{1}{N} {\bf x}_{\max}^T V {\bf x}_{\max} 
\ee 
The present calculation allows to obtain the most probable value of this overlap, conditioned
on the value $\lambda=\lambda_{\rm max}(M)$ for $\lambda \geq \lambda_{\rm typ}$.

Let us go back to the saddle point solution in Section \ref{subsec:replicaSP}. 
For the RS solution we found that the overlap at the saddle point was
given by the second equation in \eqref{uRS}, which using \eqref{defhatchic} leads to 
\be  
u = \frac{\beta T}{2}  \int dv \rho(v) v [ \frac{1}{ \kappa -  v} + \frac{ \frac{2 J^2}{\beta T}  q}{(\kappa  - v )
(\kappa -  v - \frac{2 n J^2}{\beta T}  q)} ]
\ee 
Performing the $T \to 0$, $n \to 0$ limit with $s = n/T$ fixed and using $q \to 1$, we obtain, in the
delocalized regime
\be 
u = \int dv \tilde \rho_s(v) \, v   \quad , \quad \tilde \rho_s(v)= J^2 \frac{\rho(v)}{ (\kappa - v  )(\kappa -  v - 2 s J^2/\beta)}  
\ee
i.e. $u$ is the first moment of the density $\tilde \rho_s(v)$, which by definition is normalized to unity,
see \eqref{cond1}. 
As we now discuss it is natural to
identify this density with the occupation measure.

Note that from the definition \eqref{deff} of $f(z)$, the equation $f(z_+)=f(z_-)$ with $z_+> z_-$ implies upon substracting and rearranging
\be 
1 = J^2 \int dv \frac{\rho(v)}{ (z_+ - v  )(z_- -  v)}  
\ee 
Thus, introducing the occupation measure defined in \eqref{defoccupation} 
by conditioning on the value $\lambda=\lambda_{\rm max}$, 
it is natural to surmise that in the delocalized regime it reads 
\be 
\rho_\lambda(v) = J^2  \frac{\rho(v)}{ (z_+(\lambda) - v  )(z_-(\lambda) -  v)}  
\ee 
where we recall that $z_+(\lambda)>z_-(\lambda) >v_e$ are the two
roots of $\lambda=f(z)$. One has of course the identity $\rho_\lambda(v)=\tilde \rho_s(v)$, where
$\lambda$ and $s$ are in correspondence through the Legendre transform.
\\

{\bf Remark}. Although we will not detail it here, this conjecture can be argued by
introducing $u_g= \frac{1}{N} {\bf x}^T g(V) {\bf x}$ for test
functions $g(V)$, and considering the generating function
with a infinitesimal source $j \to 0$
\be  \label{ZNT3} 
Z_N(T,j) = \int_{{\bf x} \in \Omega_N}  d {\bf x} \, \delta( {\bf x}^\dagger {\bf x} - N) \, 
e^{ \frac{1}{T} {\bf x}^\dagger (J H + V) {\bf x}  + \frac{j}{N} {\bf x}^\dagger V {\bf x} }
\simeq e^{ \frac{1}{T} \lambda_{\rm max}(M)} e^{ \frac{j}{N} {\bf x}_{\max}^T V {\bf x}_{\max} }
\ee 
Repeating the same calculation, the replica saddle point equations are only infinitesimally modified, and they lead to the saddle point value $u_g = \int dv \tilde \rho_s(v) g(v)$,
which can be identified with the conditional expectation $\langle \frac{1}{N} {\bf x}_{\max}^T g(V) {\bf x}_{\max}  \rangle$ (conditioned to $\lambda_{\rm max}=\lambda=\lambda(s))$.

%
%
%
%
%

%

\section{Quadratic optimization in presence of a random field}

Extending the study of \cite{TrivializationUs2014} we now consider 
the same quadratic optimization problem in presence of an independent Gaussian random field.
We want to study the large deviations of the random variable 
\be  
\lambda_{\sigma} = \max_{{\bf x} \in \Omega_N} ( {\bf x}^\dagger (J H + V) {\bf x}  +   {\bf h}^\dagger {\bf x}  + {\bf x}^\dagger {\bf h}  ) 
\ee 
where ${\bf h}$ is a real/complex $N$ component Gaussian random field 
with variance $\langle h_i^* h_j \rangle= \beta \sigma^2 \delta_{ij}$. 
Since we will always consider expectation values both over $H$ and the random field, 
we use the same bracket symbol. For $\sigma=0$ one recovers the case
studied above and in this section we use the notations $\lambda_0 = \lambda_{\rm max}(M)$,
$\lambda^{\rm typ}_0 = \lambda_{\rm typ}$.
One introduces the partition sum
\be  \label{ZNTRF} 
Z_N(T) = \int_{{\bf x} \in \Omega_N}  d {\bf x} \, \delta( {\bf x}^\dagger {\bf x} - N) \, 
e^{ \frac{1}{T} {\bf x}^\dagger (J H + V) {\bf x}  + \frac{1}{T}  ({\bf h}^\dagger {\bf x}  + {\bf x}^\dagger {\bf h}) }
\quad , \quad \Omega_N = \begin{cases}  \mathbb{R}^N (\beta=1) \\
 \mathbb{C}^N (\beta=2) \end{cases} 
\ee 
One can write its moments $\langle Z(T)^n \rangle$ as in Section \ref{sec:replica}.
Using that
\be 
\langle (\sum_a {\bf h}^\dagger  {\bf x}_a  + {\bf x}_a^\dagger {\bf h})^2  \rangle 
= 2 \sigma^2 \sum_{a,b}  ({\bf x}_a^\dagger {\bf x}_b + {\bf x}_b^\dagger {\bf x}_a)
\ee 
we see that in the exponential in Eq. \eqref{moment1} there is now the additional
term $\frac{2 \sigma^2}{T^2} N \sum_{ab} Q_{ab}$. Thus the action
\eqref{SQ} has the additional term $\delta S=\frac{2 \sigma^2}{T^2} \sum_{ab} Q_{ab}$. Hence the action
\eqref{action} becomes
\be \label{action2} 
S[Q,\chi,u,w]  =  \frac{i}{2} \sum_{ab} \chi_{ab} Q_{ab} + 
( \frac{i}{2} w_{a} + \frac{1}{T}) u_{a} + \frac{J^2}{\beta T^2}  {\rm tr} Q^2 + \frac{2 \sigma^2}{T^2} \sum_{ab} Q_{ab}
- \frac{\beta}{2} \int dv \rho(v) \, {\rm tr} \log (i \chi + i v w) 
\ee 
Only one of the SP equations in \eqref{4SP} is changed and it becomes
\be 
 i \chi_{a \neq b} = - \frac{4 J^2}{\beta T^2} Q^*_{a \neq b} - \frac{4 \sigma^2}{T^2} 
\ee 

Proceeding with the RS saddle-point, we see that \eqref{2eq} is unchanged, but
\eqref{uRS} becomes
\be \label{uRSRF} 
i \chi = - \frac{4}{T^2} ( \frac{J^2 q}{\beta} + \sigma^2) 
\ee 

Eliminating $\chi$ and using $i \chi_c =  \frac{2}{T} \kappa$, we obtain the system
\bea  \label{system1new} 
&& q = (J^2  q  + \beta \sigma^2) \int dv  \frac{\rho(v) }{(\kappa - v  )
(\kappa - v - 2 n (J^2 q+ \beta \sigma^2)/(\beta T))} \\
&& 1 - q = \frac{\beta T}{2} \int dv \frac{\rho(v) }{\kappa - v}  \label{system2new} 
\eea 
and the action \eqref{actionT} becomes
\be \label{actionTnew} 
 \frac{1}{n}  S[q,\kappa]  =   \frac{\kappa}{T} +  \frac{J^2}{\beta T^2}   (1 -2 q - (n-1) q^2)  
- \frac{\beta}{2 n} \int dv \rho(v) \left( (n-1) \log(\kappa - v) + \log(\kappa - v -  \frac{2 n}{\beta T} (J^2 q+ \beta \sigma^2) ) \right) 
\ee 
Note that, again, if we take derivatives w.r.t. $q$ and $\kappa$,
we obtain two equations (identical to \eqref{twoeq} with the change $J^2 q \to J^2 q + \beta \sigma^2$ in the
denominators) which are two independent linear combinations of the SP equations 
\eqref{system1new}, \eqref{system2new}.
\\

\noindent
{\bf Absence of deformation $V=0$}. Let us consider the case of pure GOE/GUE ensemble $V=0$, i.e. $\rho(v)=\delta(v)$. 
The system \eqref{system1new}-\eqref{system2new} leads to 
\be 
1 - q = \frac{\beta T}{2 \kappa}  \quad , \quad 
q = \frac{(J^2 q+\beta \sigma^2)}{\kappa (\kappa  - 2 n (J^2 q+\beta \sigma^2)/(\beta T))} 
\ee 
Eliminating $\kappa$ one finds the equation which determines $q$
\be  
(J^2 q + \beta \sigma^2) (1-q) (1 + (n-1) q) = \frac{\beta^2 T^2}{4} q
\ee  
which coincides with Eq. (29) in \cite{TrivializationUs2014} (we must change our $T \to 2 T$ and set $\beta=1$
for GOE). For $n=0$ (and $n$ near zero) as discussed there, the transition at $T=T_c = \frac{2 J}{\beta}$ in zero field disappears
in presence of a random field. There are three real roots, but only a single physical one with $0<q=q(T)<1$ for all $T>0$
(the second root is negative and the third one is larger than unity). The physical root increases
from $q=q_0(T)$ to $q=1$ as $\sigma^2/J^2$ increases monotonically from $0$ to $+\infty$, where $q_0(T)=\max(0,1-T/T_c)$
is the zero field spin glass order parameter.  
\\

To compute the CGF \eqref{LD0RF}, let us now set $n = T s$, $s \geq 0$, and take simultaneously the $n=0$ and $T=0$ limit at fixed $s$.
The SP equations \eqref{system1new}  show that $q \to 1$, more
precisely $1 - q = T q_1 + O(T^2)$  and 
\be  \label{system1new2} 
 1 = (J^2    + \beta \sigma^2) \int dv  \frac{\rho(v) }{(\kappa - v  )
(\kappa - v - 2 s (J^2 + \beta \sigma^2)/\beta)} 
\ee
and $q_1= \frac{\beta}{2} \int dv \frac{\rho(v) }{\kappa - v} $.
It determines $\kappa= \kappa(s)$ as a function of $s$. 
This is exactly the same equation as in zero field but with $J^2 \to J^2 + \beta \sigma^2$. 
Hence we already know the solutions. We introduce again the function $f(z)$ which is modified 
accordingly and becomes
\be 
f_\sigma(z) = z + (J^2 + \beta \sigma^2) \int dv \rho(v) \frac{1}{z-v} = f_0(z)|_{J^2 \to J^2 + \beta \sigma^2}
\ee 
Using that, for $z_+>z_-$
\be 
f_\sigma(z_+)=f_\sigma(z_-) \quad \Leftrightarrow \quad 
 1 = (J^2    + \beta \sigma^2) \int dv  \frac{\rho(v) }{(z_+ - v  )
(z_- - v) } 
\ee
one easily checks that the equation \eqref{system1new2} is equivalent to
\be \label{ff} 
f_\sigma(z_+)= f_\sigma(z_-) \quad , \quad \kappa=z_+  \quad , \quad \kappa - 2 s (J^2 + \beta \sigma^2)/\beta=z_-
\ee 
and one has $z_+-z_-= 2 s (J^2 + \beta \sigma^2)/\beta$.
The typical case $s=0$ corresponds to $z_+=z_-=z^*$ and $f_\sigma'(z^*)=0$. 
Clearly the roots $z_+$ and $z_-$ are obtained from the case $\sigma=0$ by the
simple substitution $J^2 \to J^2 + \beta \sigma^2$. Hence the analog of the 
delocalized phase now corresponds to 
\be
\tilde J^2 := J^2 + \beta \sigma^2 > J_c^2 
\ee 
where $J_c$ is the zero field transition coupling. We now discuss this phase.
\\

{\bf Analog of the delocalized phase, $\tilde J> J_c$}. 

{\it Delocalized regime}.
Within that phase there is an analog of a delocalized regime
$0< s <s^*$ where \eqref{ff} holds with $z_+>z_->v_e$. At the boundary of the
delocalized regime $s=s^*$ one has $z_-=v_e$. 
The value of
$s^*$ is obtained from the zero field case from the substitution $s^*=s^*(J)|_{J^2 \to J^2 + \beta \sigma^2}$, with $s^*>0$
within the delocalized phase. 
Beyond that is the localized
regime (see below).

From \eqref{actionTnew} and \eqref{PhiLim} we obtain
within the delocalized regime, i.e for $J^2+\beta \sigma^2>  J_c$ and $0 \leq s \leq s^*$,  the scaled CGF in presence of the random field as
\be \label{actionT0new} 
\phi_\sigma(s) =  \phi_\sigma(s, \kappa(s)) \quad , \quad 
 \phi(s, \kappa) :=
  \kappa s - \frac{J^2 s^2}{\beta}  
- \frac{\beta}{2} \int dv \rho(v)  \log\left( \frac{\kappa - v - 2 s (J^2+  \beta \sigma^2)/\beta}{\kappa - v} \right)  
\ee
where $\kappa=\kappa(s)$ is the solution of \eqref{system1new2}, which is
also the solution of \eqref{ff}. Again one has $\partial_{\kappa} \phi(s, \kappa)|_{\kappa=\kappa(s)} = 0$.
Then we see that 
\be 
\phi_\sigma(s) = \sigma^2 s^2 + \tilde \phi_0(s) \quad , \quad 
\tilde \phi_0(s) = \phi_0(s) |_{J^2 \to J^2+  \beta \sigma^2} 
\ee 
where $\phi_0(s)$ is the scaled CGF for $\sigma=0$ (computed in the previous sections
and denoted $\phi(s)$ there). 

We can now obtain ${\cal L}_\sigma(\lambda)$, the rate function in presence of the random
field in \eqref{LD0RF}. It is obtained by the Legendre inversion
\be 
\beta {\cal L}(\lambda) = \max_{s \geq 0} (\lambda s - \phi_\sigma(s)) 
= \max_{s \geq 0} (\lambda s - \sigma^2 s^2 - \tilde \phi_0(s))  
\ee

Let us denote ${\cal L}_0(\lambda)$ the rate function for $\sigma=0$
(computed in the previous sections
and denoted ${\cal  L}(s)$ there). Let us further denote 
\be 
\tilde {\cal L}_0( \lambda) = {\cal L}_0(\lambda)|_{J^2 \to J^2+  \beta \sigma^2} 
\ee 

With the aim of computing ${\cal L}_\sigma(\lambda)$, let us now discuss the relation between $\lambda$ and $s$ from the Legendre transform.
One has 
\be 
\lambda = \lambda(s) = \phi_\sigma'(s) =  2 \sigma^2 s + \tilde \phi'_0(s)  
= 2 \sigma^2 s + \tilde \lambda_0(s) 
\ee
where $\tilde \lambda_0(s) = \tilde \phi'_0(s)$ and due to the substitution $J^2 \to J^2+  \beta \sigma^2$ one has
\be \label{ff2} 
\tilde \lambda_0(s) = f_\sigma(z_+)=f_\sigma(z_-) \quad , \quad 
\beta \frac{z_+-z_-}{2 (J^2 + \beta \sigma^2)} = s =\beta  \tilde {\cal L}_0'(\tilde \lambda_0)
\ee 
The typical value $\lambda_\sigma^{\rm typ}$ of $\lambda_\sigma$ corresponds to $s=0$ and is thus 
\be 
\lambda_\sigma^{\rm typ}= \phi'_\sigma(0)=\tilde \phi_0'(0) = \tilde \lambda_0(0)=
\lambda_0^{\rm typ}|_{J^2 \to J^2 + \beta \sigma^2}
\ee 

Since $\tilde {\cal L}_0( \lambda)$ is known, e.g. it is 
given by \eqref{Ldeloc} upon the substitution $J \to \tilde J$,
we can thus obtain ${\cal L}(\lambda)$ parametrically as
\bea \label{relat0} 
&& \lambda = \tilde \lambda_0 + 2 \sigma^2 \beta \tilde  {\cal L}_0'(\tilde \lambda_0) \\
&&  {\cal L}'(\lambda) =  \tilde {\cal L}_0'(\tilde \lambda_0) = s/\beta \label{relat01} 
\eea 
Note that the boundary of the delocalized regime corresponds
to  $s=s^*=s^*(J,\sigma)=s^*(\tilde J)$, which, from \eqref{ff2} and since $z_-=v_e$, corresponds to 
$\tilde \lambda_0(s^*) = f_\sigma(v_e) = 
\lambda_e|_{J^2 \to J^2 + \beta \sigma^2}$. In turns this corresponds to
\be 
\lambda = \lambda_\sigma^e = f_\sigma(v_e)  + 2 \sigma^2 s^*(\tilde J) 
\ee

Hence one has in the delocalized regime, $\lambda_\sigma^{\rm typ} < \lambda < \lambda_\sigma^e$
\bea 
&& {\cal L}_\sigma(\lambda) = \int_{\lambda_\sigma^{\rm typ}}^{\lambda} d\lambda_1 {\cal L}'(\lambda_1) 
= \int_{\tilde \lambda_0 [\lambda_\sigma^{\rm typ}]}^{ \tilde \lambda_0 [\lambda]}  d \tilde \lambda_0 
(1 + 2 \sigma^2 \beta \tilde {\cal L}_0''(\tilde \lambda_0)  ) \tilde {\cal L}_0'(\tilde \lambda_0)  \\
&& = \tilde {\cal L}_0(\tilde \lambda_0[\lambda]) + \sigma^2 \beta \tilde {\cal L}_0'(\tilde \lambda_0[\lambda])^2 \label{Lfinalfield} 
\eea 
where we denote the function $\tilde \lambda_0[\lambda]$ of $\lambda$ obtained by inversion of \eqref{relat0},
with a different bracket notation, not to 
confuse it with the function $\tilde \lambda_0(s)$ of $s$ defined above. 
Since $\tilde \lambda_0 [\lambda_\sigma^{\rm typ}]=\tilde \lambda_0(0)$
is the typical value of the problem without a field and
with the substitution $J \to \tilde J$, one has 
$\tilde {\cal L}_0(\tilde \lambda_0(0))=\tilde {\cal L}'_0(\tilde \lambda_0(0))=0$,
hence one also has ${\cal L}_\sigma(\lambda_\sigma^{\rm typ})= 
{\cal L}'_\sigma(\lambda_\sigma^{\rm typ})=0$ as required.
In addition one has $\tilde \lambda_0[\lambda_\sigma^e]= \tilde \lambda_0(s^*)$. 
\\

{\it Analog of localized regime}.
Let us consider now, still within the analog of the delocalized phase, the region $s \geq s^*$.
Again $\kappa$ freezes at its boundary value (which is $s$ dependent)
\be \label{frozenchisigma} 
 \kappa = v_e + 2 s (J^2 +\beta  \sigma^2) /\beta
\ee
Then \eqref{system1new2} 
does not hold anymore and one can use only the SP equation
$\partial_q S[q,\kappa]=0$ in \eqref{actionTnew},
which again, upon inserting $s= n T$, leads to $1-q = O(T)$. 
Injecting this value of $q$ together with $\kappa$ from \eqref{frozenchisigma} 
into the expression for the action \eqref{actionTnew} and obtain using 
\eqref{PhiLim} the scaled CGF $\phi_\sigma(s)$ for $J \geq J_c$ and $s \geq s^*$, 
as 
\be  \label{Philocreg2} 
\phi_\sigma(s)  = \phi_\sigma(s,v_e + 2 s \tilde J^2/\beta) = 
  v_e s + \frac{J^2 s^2}{\beta}  + 2 s^2 \sigma^2
- \frac{\beta}{2} \int dv \rho(v)  \log\left( \frac{v_e - v }{v_e - v + 2 s (J^2+ \beta \sigma^2)/\beta} \right)  
\ee 
where $\phi(s,\kappa)$ is given by the same formula as 
\eqref{actionT0new}. Hence the relation
\be 
\phi_\sigma(s) = \sigma^2 s^2 + \tilde \phi_0(s) \quad , \quad 
\tilde \phi_0(s) = \phi_0(s) |_{J^2 \to J^2+  \beta \sigma^2} 
\ee 
still holds in the localized regime. As a consequence
the relations \eqref{relat0}, \eqref{relat01} and 
\eqref{Lfinalfield}, which determine ${\cal L}_\sigma(\lambda)$ 
for $\lambda > \lambda_\sigma^e$, also hold there. 
\\

\noindent
{\bf Case $V=0$}. As a check, let us consider GOE/GUE matrices without perturbation. 
There is only a delocalized regime. Performing the substitution $J^2 \to J^2+  \beta \sigma^2$ 
one has from \eqref{standard}  for $\lambda \geq \lambda_\sigma^{\rm typ}=2 \tilde J$ with $\tilde J= \sqrt{J^2 + \beta \sigma^2}$,

\be  
\tilde {\cal L}_0'(\lambda) = \frac{1}{2 \tilde J^2}  \sqrt{\lambda^2 - 4 \tilde J^2} \quad , \quad 
\tilde {\cal L}_0(\lambda) = \frac{\lambda}{4 \tilde J^2}  \sqrt{\lambda^2 - 4 \tilde J^2} - \log( \frac{\lambda + \sqrt{\lambda^2 - 4 \tilde J^2}}{2 \tilde J}) 
\ee 
Hence we have
\bea \label{lala0} 
&& \lambda = \tilde \lambda_0 +  \frac{ \sigma^2 \beta}{ \tilde J^2}  \sqrt{\tilde \lambda_0^2 - 4 \tilde J^2}  \quad , 
\quad 
 {\cal L}'(\lambda) =  \frac{1}{2 \tilde J^2}  \sqrt{\tilde \lambda_0^2 - 4 \tilde J^2} 
\eea 
We find, since $\lambda \geq \tilde \lambda_0$ 
%
To make contact with \cite{TrivializationUs2014} let us 
use the notation $\Gamma= \frac{\sigma^2 \beta}{J^2}$,
which leads to $\tilde J^2 = J^2 (1+ \Gamma)$. One can 
invert \eqref{lala0} and obtain 
\be \label{invlala0} 
\tilde \lambda_0 = \tilde \lambda_0[\lambda] = \frac{(1+ \Gamma)}{1+ 2 \Gamma}  \left( 
(1+ \Gamma) \lambda - \Gamma
\sqrt{\lambda^2 - 4 J^2 \frac{1 + 2 \Gamma}{1+ \Gamma} } \right)
\ee 
Computing the two terms in \eqref{Lfinalfield} 
\bea 
&& \tilde {\cal L}_0(\tilde \lambda_0) = \frac{\tilde \lambda_0}{4 J^2 (1+\Gamma)}  \sqrt{\tilde \lambda_0^2 - 4 J^2 (1+ \Gamma)} - \log( \frac{\tilde \lambda_0 + \sqrt{\tilde \lambda_0^2 - 4 J^2 (1+ \Gamma)}}{2 J \sqrt{1+ \Gamma}}) 
\eea 

\bea 
&& \sigma^2 \beta {\cal L}_0'(\tilde \lambda_0(\lambda))^2 
 = \frac{\Gamma}{4 J^2 (1 + \Gamma)^2}  ( \tilde \lambda_0^2 - 4 J^2(1+ \Gamma)  ) 
\eea 
and inserting \eqref{invlala0}, the final result reads
(we set here $J=1$ which amounts to express $\lambda$ in units of $J$) 
\be 
{\cal L}(\lambda)=
-\frac{\Gamma  \lambda ^2}{4(1 + 2  \Gamma)}+\frac{(\Gamma +1) \lambda 
   \sqrt{\lambda ^2- \lambda_c^2}}{4( 1 + 2 \Gamma)}-\log
   \left(\frac{\sqrt{\Gamma +1}
   \left(\lambda +\sqrt{\lambda
   ^2-\lambda_c^2}\right)}{2 (2
   \Gamma +1)}\right) \quad , \quad \lambda_c^2 = 4 \frac{1+ 2 \Gamma}{1 + \Gamma} 
   \ee 
This coincides with the result in Eq.(43) of \cite{TrivializationUs2014}
with the correspondence ${\cal E}= - \lambda/2$. The rate function
vanishes as required for the typical value 
$\lambda=2 \sqrt{1+ \Gamma}$. 

\bigskip

{\bf Acknowledgments}. 
I thank Y. Fyodorov, A. Krajenbrink and B. Lacroix-A-Chez-Toine for discussions and 
collaborations on related topics.
I acknowledge support from ANR Grant No. ANR- 23-CE30-0020-0.

\bigskip

\bigskip

\appendix

\begin{center}
\begin{large}
{\bf Appendix} 
\end{large}
\end{center}

\section{Derivation of RS equations} \label{app:derivation} 

Note that the variation of this RS action gives 
\bea 
&& \frac{\delta}{\delta i \chi} \quad \Rightarrow \quad 1 - q + n q = \int dv \rho(v) \frac{1}{i \chi_c - \beta v + n i \chi}  \\
&& \frac{\delta}{\delta q} \quad \Rightarrow \quad i \chi (n-1) + \beta^2 J^2 q (n-1) = 0 \\
&& \frac{\delta}{\delta i \chi_c} \quad \Rightarrow \quad 1 -  \int dv \rho(v) \frac{1}{i \chi_c - \beta v} 
+ \frac{1}{n}  \int dv \rho(v) \left( \frac{1}{i \chi_c - \beta v } - \frac{1}{i \chi_c - \beta v + n i \chi} 
\right) = 0 
\eea 
The second equation, for $n \neq 1$ is equivalent to $i \chi = - \beta^2 J^2 q$. The first equation can be obtained
from combining the two equations in \eqref{2eq}. The last equation can also be obtained
from another combination of the two equations in \eqref{2eq}. Hence the SP equations can be 
obtained after substitution (which will be useful in the following). 

\section{Convexity of $\phi(s)$} \label{app:convex} 

One can compute $\phi''(s)$ to check the convexity. In the delocalized regime 
starting from \eqref{Phideloc0} and using that $\partial_{z} \phi(s, z) |_{z={\sf z}_+(s)} = 0$
one has
\be 
\phi''(s) = \partial_s^2 \phi(s,z)|_{z={\sf z}_+(s)}   + z'_+(s) \partial_s \partial_z \phi(s,z)|_{z={\sf z}_+(s)} 
\ee 
One can rewrite $f({\sf z}_+(s))=f({\sf z}_-(s)={\sf z}_+(s) - \frac{2 J^2 s}{\beta})$ in the form
\be 
\int_{{\sf z}_+(s) - \frac{2 J^2 s}{\beta}}^{{\sf z}_+(s)} dz f'(z) = 0 
\ee 
and take derivatives which leads to 
\be 
\frac{d{\sf z}_+}{ds} = - \frac{2 J^2}{\beta} \frac{f'(z_-)}{f'(z_+)-f'(z_-)}
\ee 
where here and below we simply denote ${\sf z}_\pm(s)=z_\pm$ (which
we recall are also the two roots of $\lambda=f(z)$ where $\lambda$ and $s$
are associated by the Legendre transform).
Using that 
\bea 
&& \partial_s \phi(s,z) = f(z - 2 s J^2/\beta) \quad , \quad  \partial_z \partial_s \phi(s,z)|_{z={\sf z}_+(s)}  
= f'(z_-) \\
&& \partial^2_s \phi(s,z)|_{z={\sf z}_+(s)} = - 2 \frac{J^2}{\beta} f'(z_-) 
\eea 
the calculation simplifies and one obtains 
\be 
\phi''(s) = - \frac{2 J^2}{\beta} \frac{f'(z_+) f'(z_-)}{f'(z_+)-f'(z_-)} >0 
\ee 
since in the delocalized regime $f'(z_-)<0$ and $f'(z_+)>0$, see Fig.
\ref{fig:Sketch}.
\\

In the localized regime one has from \eqref{104},  $\phi'(s)=f(v_e + 2 \frac{J^2}{\beta} s)$.
Since $f'(z_+)>0$ in that regime, one has $\phi''(s)>0$. 
\\

In the case of $V=0$, i.e. pure GOE/GUE $\rho(v)=\delta(v)$, where 
there is only a delocalized phase and a delocalized regime one recovers,
using that $z_\pm= \frac{J^2}{\beta} (\pm s +  \sqrt{s^2 + \frac{\beta^2}{J^2} } )$ 
and $f(z)=z+J^2/z$, for $s>0$
\be 
\phi''(s) = \frac{2 s J^2}{\beta \sqrt{s^2 + \beta^2/J^2}} >0
\ee

\section{More details on the interpolation in the critical regime}
\label{app:interpolation}

Let us consider the localized side $J<J_c$. Near the transition $0< f'(v_e)=1-\frac{J^2}{J_c^2} \ll 1$, while $f''(v_e)$ and the higher derivatives are of order one.
We need to solve $\lambda=f(z_+)$ for $\lambda$ near $\lambda_e$ and $z_+$ near $v_e$.
Let us denote 
\be 
\delta \lambda = \lambda-\lambda_e \quad , \quad z_\pm=v_e+\delta z_\pm
\ee
Keeping only the relevant terms we need to solve
\be \label{tt} 
\delta \lambda = f'(v_e) \delta z + \frac{1}{2} f''(v_e) \delta z^2 
\ee 
This leads to
\be 
\delta z_+ =  \sqrt{\epsilon^2 + 2 \frac{\delta \lambda}{f''(v_e)}} - \epsilon \quad , \quad \epsilon= \frac{f'(v_e)}{f''(v_e)} 
\ee 
Using \eqref{Lprime2} one finds
\be 
{\cal L}(\lambda) = \frac{1}{2 J^2} \bigg(  \frac{f''(v_e)}{3} \left(  (\epsilon^2 + 2 \frac{\delta \lambda}{f''(v_e)} )^{3/2} - \epsilon^3 \right) 
- \epsilon \,  \delta \lambda \bigg) 
\ee 
which interpolates between 
\bea  
&& {\cal L}(\lambda)  \simeq \frac{\delta \lambda^2}{4 J^2 f'(v_e)} \quad , \quad \delta \lambda \ll \frac{f'(v_e)^2}{f''(v_e)} \\
&& {\cal L}(\lambda)  \simeq  \frac{1}{3 J^2} \sqrt{\frac{2}{f''(v_e)}} \delta \lambda^{3/2} \quad , \quad \delta \lambda \gg \frac{f'(v_e)^2}{f''(v_e)}
\eea
Note that the above is true only for $\epsilon \ll 1$ and $\delta \lambda = O(\epsilon^2)$.
When $f'(v_e)$ is not small only the first regime holds as in \eqref{Lloc}. 
\\

Let us consider now the delocalized side $J>J_c$, very near the transition. Now we have $f'(v_e)<0$
and very small. One has to be careful that on that side there are two regimes, the delocalized 
one for $\lambda_{\rm typ} < \lambda < \lambda_e$ and the localized one for $\lambda>\lambda_e$. 
Consider first the delocalized regime.
One still has to solve \eqref{tt} but there are now two roots $z_\pm=v_e \pm \delta z_\pm$ close
to $v_e$. One has, with $\delta \lambda= \lambda - \lambda_e$,
\be 
\delta z_\pm = |\epsilon| \pm \sqrt{\epsilon^2 + 2 \frac{\delta \lambda}{f''(v_e)}}   \quad , \quad \epsilon= \frac{f'(v_e)}{f''(v_e)} <0 
\ee 
Now we have to remember that $\lambda_e$ is not the typical value, and instead one has
$\lambda_{\rm typ}= f(z^*)$ where $f'(z^*)=0$. One finds that 
$z^*-\lambda_e=-\epsilon$ and $\lambda_{\rm typ}-\lambda_e \simeq - f'(v_e) \epsilon + \frac{1}{2} f''(v_e) \epsilon^2
= - \frac{f'(v_e)^2}{2 f''(v_e)}$. Hence $\delta \lambda= \lambda - \lambda_{\rm typ} - \frac{f'(v_e)^2}{2 f''(v_e)}$.
Using that ${\cal L}'(\lambda)=(z_+-z_-)/(2 J^2) =(\delta z_+-\delta z_-)/(2 J^2)$ we obtain by integration
that for $\lambda_{\rm typ} < \lambda < \lambda_e$
\be 
{\cal L}(\lambda) =  \int_{\lambda_{\rm typ}}^{\lambda} dx {\cal L}'(x)
=  \int_{\lambda_{\rm typ}}^{\lambda} dx \frac{\delta z_+-\delta z_-}{2 J^2}
= \frac{2}{3 J^2} \sqrt{\frac{2}{f''(v_e)}} (\lambda - \lambda_{\rm typ})^{3/2} 
\ee 
One has
\be  
{\cal L}(\lambda_e) = \frac{f''(v_e)}{3 J^2} |\epsilon|^3
\ee   
Now for $\lambda>\lambda_e$, i.e. in the localized regime, 
one has again ${\cal L}'(\lambda)=(z_+-v_e)/(2 J^2) =\delta z_+/(2 J^2)$ and
\bea 
 {\cal L}(\lambda) &=&  {\cal L}(\lambda_e)  + \int_{\lambda_{e}}^{\lambda} dx \frac{\delta z_+(x)}{2 J^2} 
\\
&=& 
\frac{f''(v_e)}{3 J^2} |\epsilon|^3 + \frac{1}{2 J^2} \bigg(  \frac{f''(v_e)}{3} \left(  (\epsilon^2 + 2 \frac{\lambda-\lambda_e}{f''(v_e)} )^{3/2} - |\epsilon|^3 \right) 
+ |\epsilon| \,  ( \lambda-\lambda_e)  \bigg) \\
& = & \frac{1}{2 J^2} \bigg(  \frac{f''(v_e)}{3}  (2 \frac{\lambda-\lambda_{\rm typ}}{f''(v_e)} )^{3/2}  
+ |\epsilon| \,  ( \lambda-\lambda_{\rm typ})  - \frac{1}{6} f''(v_e) |\epsilon|^3 \bigg) \\
& = & \frac{1}{3 J^2} \sqrt{\frac{2}{f''(v_e)}} (\lambda-\lambda_{\rm typ})^{3/2}  
+ \frac{1}{2 J^2} |\epsilon| (\lambda-\lambda_{\rm typ})
 - \frac{1}{12 J^2} f''(v_e) |\epsilon|^3 
\eea

\section{Comparison with Ref. \cite{McKenna2021} (McKenna)} \label{app:McKenna} 

In \cite{McKenna2021} McKenna provides a rigorous computation of the
large deviation function for the deformed GOE/GUE ensemble.
That work corresponds to $J=1$, our $\lambda$ is called $x$, 
our $\rho(v) dv$ is called $\mu_D$ with an upper edge at
$v_e=r(\mu_D)$. His parameter $x_c$ is our $\lambda_e=f(v_e)$. The main result is theorem 2.7 on page 5. 
It is expressed as a variational problem w.r.t. the parameter denoted $\theta$, and one denotes
$\theta_x=\frac{\beta}{2} \tilde \theta_x$ the value at the optimum. In Remark 2.13 some
simpler formula are given. 

We would like to identify $I_\beta(x)= \beta {\cal L}(\lambda=x)$. We did not 
fully establish it, but we at least we show some consistency. 

{\bf Localized regime $x>x_c$}.  The result of \cite{McKenna2021} appears simpler in the case $x>x_c$ (i.e. $\lambda>\lambda_e$),
which we call the localized regime. In that case the optimal is showed to be $\tilde \theta=x-v_e$.
From the un-numbered equation above (2.14) in \cite{McKenna2021},
taking a derivative we find
\be \label{99}
\frac{2}{\beta} I_\beta'(x) = x- v_e - G_M(x) \quad , \quad G_M(x) = G_V(x - G_M(x)) 
\ee 
We would like to identify $I'(x)=s$ where $s$ is our variable $s$. If we do that we obtain
\be 
x- v_e - G_M(x) = 2 s/\beta 
\ee 
Using the second relation in \eqref{99} it implies that $G_M(x)=G_V(v_e + \frac{2 s}{\beta})$,
which, inserted in the first relation of \eqref{99}  leads to 
\be 
x=   v_e + \frac{2 s}{\beta}  + G_V(v_e + \frac{2 s}{\beta} ) =
v_e + \frac{2 s}{\beta} +  \int dv  \frac{\rho(v)} {v_e - v + 2 s/\beta} = \lambda
\ee 
which is exactly our equation \eqref{104}. So it is consistent, at the
level of the derivatives, that is $I'(x)=s$ and $\phi'(s)=x$.
Once $I'(x)=x-v_e - G_M(x)$ is established
the integration constant can in principle be fixed 
by using that the rate function should vanish at $x=\lambda=\lambda_{\rm typ}=v_e$. 
\\


{\bf Delocalized regime $x<x_c$}. In the delocalized regime the optimal $\theta$ of \cite{McKenna2021},
$\theta_x=\frac{\beta}{2} \tilde \theta_x$ is shown to obey
\be \label{eqmac} 
\tilde \theta + K(\tilde \theta) = x \quad  \Leftrightarrow \quad \tilde \theta = G_V(x-\tilde \theta) = \int dv \frac{\rho(v)}{x-\tilde \theta-v} 
\ee 
If we compare with our equation 
\be \label{eqeq2} 
\lambda =  f(z) = z
+ J^2  \int dv  \frac{\rho(v)} {z - v} 
\ee 
we see that it is identical provided we identify $z= x - \tilde \theta$.
Since for $x>x_c$ it is said in \cite{McKenna2021} that $\tilde \theta=x-v_e$ 
then we should identify 
\be 
x - \tilde \theta = z_- 
\ee 
We were not able to compare the results beyond that, nor to understand
why it is claimed there that no general explicit form may exist.
\\

{\bf The example of a semi-circle deformation}. For a more explicit comparison, 
let us consider the case where $\rho(v)$ is itself a semi-circle
\be 
\rho(v) = \frac{1}{2 \pi \sigma^2} \sqrt{4 \sigma^2 - v^2} 
\ee 
with $\sigma>0$, and we set $J=1$. The upper edge is at $v=v_e=2 \sigma$. The equation \eqref{eqf} reads, for $z \geq v_e$
\be \label{mastereq} 
\lambda = f(z) = z + \int dv \frac{\rho(v)}{z-v} = z + G_V(z) = z + \frac{1}{2 \sigma^2} (z - \sqrt{z^2- 4 \sigma^2}) 
\ee 
Solving $f'(z^*)=0$ we find the position of the minimum, $z^*$ and the typical value
\be 
z^* = \frac{1 + 2 \sigma^2}{\sqrt{1+ \sigma^2}}  \quad , \quad \lambda_{\rm typ}= f(z^*) = 2 \sqrt{1+\sigma^2}
\ee 
Since one finds that $z^*$ is always larger than $v_2=2 \sigma$ it implies that there is no localized
phase (in agreement with the general theory since $\int dv \frac{\rho(v)}{(v_e-v)^2}=+\infty$). 

Consider now $\lambda>\lambda_{\rm typ}$. The value $\lambda_e$ which separates the
delocalized regime, for $\lambda<\lambda_e$ and the localized regime $\lambda>\lambda_e$
is 
\be 
\lambda_e= f(v_e) = f(2 \sigma) = 2 \sigma + \frac{1}{\sigma} 
\ee 
In the delocalized regime the two roots of \eqref{mastereq} are
\be 
z_\pm = \frac{\lambda (1+ 2 \sigma^2) \pm \sqrt{\lambda^2 - 4 ( 1+ \sigma^2)}}{2(1 +  \sigma^2)} 
\ee 
Precisely at $\lambda=\lambda_e$ the smaller root reaches $z_-=v_e =2 \sigma$. 

In the delocalized regime $2 \sigma < \lambda < \lambda_e$ let us use (for $J=1$) 
\be 
{\cal L}'(\lambda) = \frac{1}{2} (z_+-z_-) = 
 \frac{ \sqrt{\lambda^2 - 4 ( 1+ \sigma^2)}}{2(1 +  \sigma^2)} 
\ee 
Upon integration one finds, for $2 \sigma < \lambda < \lambda_e$
\be
{\cal L}(\lambda) = \int_{\lambda_{\rm typ}}^{\lambda} dx {\cal L}'(x) 
= \frac{\lambda \sqrt{\lambda^2 -4 (1+ \sigma^2)}}{4
   \left(1+ \sigma^2\right)} + \log \frac{2 \sqrt{1+\sigma^2}}{\lambda + \sqrt{\lambda^2 - 4 (1+\sigma^2)} }
\ee

In the localized regime one has 
\be 
{\cal L}'(\lambda) = \frac{1}{2} (z_+-v_e) = 
 \frac{\lambda (1+ 2 \sigma^2) \pm \sqrt{\lambda^2 - 4 ( 1+ \sigma^2)}}{4(1 +  \sigma^2)} - \sigma
 \ee 
and
\be
{\cal L}(\lambda) = {\cal L}(\lambda_e)  + \int_{\lambda_{e}}^{\lambda} dx {\cal L}'(x) 
= \frac{\lambda  \sqrt{\lambda ^2-4
   \sigma ^2-4}-\lambda ^2}{8
   \left(\sigma
   ^2+1\right)}+\frac{1}{2} \log
   \left(\frac{2 \sigma
   }{\sqrt{\lambda ^2-4 \sigma
   ^2-4}+\lambda
   }\right)+\frac{1}{4} (\lambda -2
   \sigma )^2+\frac{1}{2} 
\ee
where
\be 
{\cal L}(\lambda_e)  = \frac{1}{4} (\frac{1}{\sigma^2} + \frac{1}{1+\sigma^2} - 2 \log(1+\frac{1}{\sigma^2}) ) 
\ee 
This coincides exactly with the result in Section 2.3  of \cite{McKenna2021}.

\end{document}